\documentclass{article}
\usepackage{graphicx}  
\usepackage{amsmath}   
\usepackage[compress]{cite}
\usepackage{amssymb}   
\usepackage{bm} 
\usepackage{dcolumn}
\usepackage{color}
\usepackage{mathrsfs}
\usepackage{amsfonts}
\RequirePackage[colorlinks,citecolor=blue,urlcolor=magenta,linkcolor=blue]{hyperref}
\def\eq#1{{Eq.~(\ref{#1})}}
\def\eqs#1{{Eqs.~(\ref{#1})}}
\def\sect#1{{Sec.~(\ref{#1})}}

\def\EH{Einstein-Hilbert }
\def\LL{Lanczos-Lovelock }
\def\gr{general relativity }
\def\RN{Reissner-Nordstr\"{o}m }
\title{Aspects of Neutrino Oscillation in Alternative Gravity Theories}
\author{Sumanta Chakraborty
\footnote{sumanta@iucaa.in}
\footnote{sumantac.physics@gmail.com}\\
{\small {IUCAA, Post Bag 4, Ganeshkhind,}}\\
{\small {Pune University Campus, Pune 411 007, India}}}
\begin{document}
  
\maketitle
\begin{abstract}
Neutrino spin and flavour oscillation in curved spacetime have been studied for the most general static spherically symmetric configuration. Having exploited the spherical symmetry we have confined ourselves to the equatorial plane in order to determine the spin and flavour oscillation frequency in this general set-up. Using the symmetry properties we have derived spin oscillation frequency for neutrino moving along a geodesic or in a circular orbit. Starting from the expression of neutrino spin oscillation frequency we have shown that even in this general context, in high energy limit the spin oscillation frequency for neutrino moving along circular orbit vanishes. We have verified previous results along this line by transforming to Schwarzschild coordinates under appropriate limit. This finally lends itself to the probability of neutrino helicity flip which turns out to be non-zero. While for neutrino flavour oscillation we have derived general results for oscillation phase, which subsequently 
have been applied to three different gravity theories. One, of them appears as low-energy approximation to 
string theory, where we have an additional field, namely, dilaton field coupled to Maxwell field tensor. This yields a realization of Reissner-Nordstr\"{o}m solution in string theory at low-energy. Next one corresponds to generalization of Schwarzschild solution by introduction of quadratic curvature terms of all possible form to the Einstein-Hilbert action. Finally, we have also discussed regular black hole solutions. In all these cases the flavour oscillation probabilities can be determined for solar neutrinos and thus can be used to put bounds on the parameters of these gravity theories. While for spin oscillation probability, we have considered two cases, Gauss-Bonnet term added to the Einstein-Hilbert action and the f(R) gravity theory. In both these cases we could impose bounds on the parameters which are consistent with previous considerations. In a nutshell, in this work we have presented both spin and flavour oscillation frequency of neutrino in most general static spherically symmetric spacetime, 
encompassing a vast class of solutions, which when applied to three such instances in alternative theories for flavour oscillation and two alternative theories for spin oscillation put bounds on the parameters of these theories. Implications are also discussed.
\end{abstract}
\newpage
\tableofcontents
\newpage 
\section{Introduction}\label{Neu:Sec:Intro}

Neutrino oscillation serves as one of the most cultivating fields in the discipline of elementary particle physics. Research along these lines have been boosted after various new experiments on solar and reactor neutrinos started yielding tantalizing results \cite{Fukuda1996,Hosaka2006,Cravens2008,Abe2011,Ahmad2002,Aharmin2005,Aharmin2008,Bellini2010}. Despite of its direct connection to the elementary particle physics neutrino oscillation has important contributions to cosmology and astrophysics as well. Even though the neutrino oscillation was put to firm experimental grounds only recently, the theoretical prediction comes much earlier from the work of Pontecorvo \cite{Pontecorvo1957} and then subsequently it was generalized for propagation within varying density medium in \cite{Mikheyev1986,Wolfenstein1978}. The oscillation we have discussed so far corresponds to neutrino flavour oscillation. There is another oscillation between various helicity states within a single flavour, known as neutrino spin 
oscillation. We could also have a mixture of both flavour and spin oscillations, which corresponds to neutrino spin-flavour oscillation.

One of the key research area in neutrino oscillation is the determination of neutrino mass and their mixing angles. This focuses on the important fact that even though the mass squared difference between various neutrino flavours and their mixing angles are well known, there is ambiguity in the determination of absolute values of neutrino masses. This leads to the famous paradigm that neutrino masses are either hierarchical or quasi-degenerate in their very nature. These results lend themselves to various important questions, which have been key research topics in recent years. These include: non-zero value for the mixing angle $\theta _{13}$, possible aspects of CP violation in neutrino oscillation and importantly the non-vanishing $1-2$ mixing angle. A large number of theoretical models exist to explain the above results. These theoretical models include, neutrino mass considered to be degenerate by some sea-saw mechanism, assumption of large mixing angle for solar as well as for atmospheric neutrino and 
use of renormalization group equation \cite{Adhikari2007,Klinkhamer2006,Schwetz2007,Cuesta2008,Akhmedov2008,Bilenky1999,Kiers1996,Sarkar1988}.

Another important place to look for effects of neutrino oscillation is astrophysics. The effects being numerous. Since neutrinos are sterile to most of the interactions, after creation they come out unimpeded and contain important information about the source. This is the main reason for setting detectors like Icecube in order to detect ultra-high energy neutrino from galactic centers and active galactic nuclei. There are also other signatures of neutrino oscillation like in the Pulsar kick mechanism devised from spin-flavour oscillation of neutrino. The behavior of neutrino in high magnetic field existing surrounding a pulsar can lead to resonant oscillation and \textit{neutrinosphere}. These \textit{neutrinospheres} can explain high proper motion of the pulsar with respect to the neighbouring stars \cite{Lambiase2005a,Lambiase2005b}. 

Neutrino oscillation was first formulated in flat spacetime which was subsequently generalized to curved spacetime \cite{Ahluwalia1996,Piriz1996,Fornengo1997,Zhang2001} and the formalism can also be used to test the equivalence principle \cite{Mann1996}. It was well known that for black hole solutions of \gr the oscillation phase along timelike geodesic yields a factor of 2 in comparison to null geodesics. The reason behind being the fact that even though the neutrino moves along null geodesic i.e. $ds^{2}=0$, the neutrino being massive satisfies the relation, $p^{2}=-m^{2}$ \cite{Bhattacharya1999,Lipkin2000,Grossman1997}. This discrepancy of factor 2 between oscillation phase along null and timelike geodesic persists for a spherically symmetric metric with $g_{tt}=g^{rr}$ \cite{Chakraborty2014CQG} and has been reproduced here for the most general of such situations. 

Along with neutrino flavour oscillation, the spin oscillation of neutrino is another important aspect. In gravitational field the spin of a particle precess and thus there is a non-zero probability of neutrino spin flip. There are two possible ways in which the spin oscillation can be studied. The first one corresponds to starting with Dirac Hamiltonian in gravitational field and then obtaining mixing terms among various chirality and thus establishing spin oscillation \cite{Casini1994,Dvornikov2005}. The other option is to work with dynamics of spinning particle in gravitational field following \cite{Papapetrou1951,Padmanabhan2010b}. The spin oscillation probability for circular motion of neutrino in Schwarzschild spacetime has been derived in \cite{Dvornikov2006} and the generalization to rotating black holes in general relativity has been done in \cite{Alavi2013}.

However all the above studies of neutrino spin and flavour oscillations have been performed in general relativity (except in \cite{Chakraborty2014CQG}). Then it is important to understand how these ideas reconcile with alternative theories of gravity since there is a general belief that \gr is only a low energy approximation of an underlying fundamental theory \cite{Buchbinder1992,Vassilevich2003}. In this spirit the \EH action for \gr gets modified by addition of higher derivative and higher order curvature terms. There exist large number of ways in which these higher order terms can be introduced in the standard \EH action. However the criteria that field equations should remain second order in the dynamical variable (otherwise some ghost fields would appear) uniquely fixes the action to be the Lanczos-Lovelock action \cite{Lanczos1932,Lovelock1971,Padmanabhan2014,Chakraborty2014PRD9012,Chakraborty2014PRD9008,Chakraborty2015sub}. Another such model explaining the above mentioned problems is obtained by 
replacing $R$, the scalar curvature in the \EH action by some arbitrary function of the scalar curvature $f(R)$. This alternative theory for describing gravitational interaction is very interesting in its on spite, for it can provide an explanation to various problems in existing general relativity as well as it confronts equally well with various tests of general relativity \cite{Bamba2008,Abdalla2005,Nojiri2011,Nojiri2003,Nojiri2008}. In this context we should also mention dilaton couplings in string inspired models, where non-zero electromagnetic fields gets coupled to dilaton \cite{Garfinkle1991,Coleman1983,Vega1988} and results in substantial modifications to \RN solution in general relativity. Also we can modify the gravity theory by introducing quadratic curvatures of various forms. This has been discussed along with possible modifications of the Schwarzschild solution in \cite{Yunes2011,Chakraborty2014PRD8902}. There also exist black hole solutions, which are regular and physical singularity is non-
existent. These solutions involve modified gravity, properly dressed additional fields, such that energy conditions leading to physical singularity are violated \cite{Hawking1973}. The first such regular black hole solution was obtained by Bardeen with magnetic charge and its relation to non-linear electrodynamics was shown by Ay\'{o}n-Beato and Garc\'{i}a \cite{ABG2000} (henceforth referred to as ABG). However the solution was \emph{non-exact}. The \emph{exact} solution to Einstein's equation with non linear electrodynamics was obtained by Ay\'{o}n-Beato and Garc\'{i}a later in \cite{ABG1998,ABG1999a,ABG1999b}. We have applied the formalism devised by us for the most general static spherically symmetric spacetime to these three black hole solutions described earlier and have studied the effect of these alternative theories to the neutrino flavour oscillation.

For neutrino spin oscillation we have considered two more alternative gravity theories both having interesting theoretical properties. The first one corresponds to vacuum solution in $f(R)$ gravity. As we have already mentioned $f(R)$ gravity is a very interesting alternative theory as it can explain all the three cosmological phases (for an alternative model see \cite{Pan2013,Pan2014}), it passes through various local tests for general relativity \cite{Bamba2008,Abdalla2005,Nojiri2011,Nojiri2003,Nojiri2008}, moreover it can explain absence of graviton Kaluza Klein modes in LHC \cite{Chakraborty2014PRD9004}. Motivated by these successes we consider vacuum solution in $f(R)$ gravity which corresponds to the Schwarzschild (Anti) de-Sitter solution. Secondly, we consider the second order \LL term with additional Maxwell field in addition to the \EH Lagrangian. This is known as the Einstein-Maxwell-Gauss-Bonnet (EMGB) gravity. This alternative theory as well posses static spherically symmetric solutions. We have 
considered the spin oscillation probability in these two alternative theories and imposed bounds on the parameters. It turns out that these bounds are consistent with previous results. 

In brief, in this work we have generalized previous works on neutrino spin and flavour oscillation to most general static spherically symmetric spacetime. To our surprise even in this general context we can make interesting predictions like, \emph{for a neutrino on a circular orbit in high energy limit the oscillation frequency vanishes}. Hence a high energy neutrino will not have any spin flip as long as it is on a circular orbit. We have also derived the oscillation frequency for geodesic motion and have used it to compute the probability of helicity flip. On the other hand we have also derived oscillation phase and oscillation length for neutrino flavour oscillation which has been used to constrain the alternative theories. In this context we have an interesting result that oscillation length depends only on the $g_{tt}$ component. It does not depend on any other metric components except for conserved energy and angular momentum. 

The paper is organized as follows: we start with a broad introduction to the spin oscillation in presence of gravitational field in \sect{Neu:Sec:SpinIntro} and then to spin oscillation frequency in \sect{Neu:Sec:SpinOsc}. After that we consider general formulation for neutrino flavour oscillation and the application to alternative theories in \sect{Neu:Sec:FlavOsc}. Finally, we provide our analysis for neutrino helicity flip in \sect{Neu:Sec:Hel} with concluding remarks in \sect{Neu:Sec:Dis}. The detailed calculations are being provided in the two appendices, Appendix \ref{Neu:AppA}, which contain all relevant calculations for both spin and flavour oscillation and in Appendix \ref{Neu:AppB}, where we have illustrated the circular orbit related issues for our general static, spherically symmetric spacetime.

\section{A Brief Introduction to Neutrino Spin Oscillation in Gravitational Field}\label{Neu:Sec:SpinIntro}

Spin angular momentum of a body gets affected by the curvature of spacetime, for example, a gyroscope orbiting around a massive body will undergo a precession of its spin. Motion of a spinning particle in gravitational field is discussed with great detail in Ref. \cite{Papapetrou1951}. Spin of a spinning particle can be presented by the use of the spin tensor $S^{\mu \nu}$ and it's momentum $p^{\mu}$ such that we can introduce the spin vector
\begin{equation}\label{Neu:Sec1:01}
S_{\rho}=\frac{1}{2m}\sqrt{-g}\epsilon _{\mu \nu \lambda \rho}p^{\mu}S^{\nu \lambda}
\end{equation}
where, we have introduced the completely antisymmetric tensor density $\epsilon _{\mu \nu \lambda \rho}$, the determinant of the metric $g=\textrm{det}(g_{\mu \nu})$ and the on-mass shell condition, $p_{\mu}p^{\mu}=-m^{2}$. For point particles, using principle of general covariance it can be shown that the evolution of spin vector $S^{\mu}$ and four-velocity $U^{\mu}$ are parallel transported along its world-line. Also the fact that spin and velocity four-vectors are orthogonal to each other remains true even in curved spacetime i.e. $U_{\mu}S^{\mu}=0$. 

However, the properties of the particle are determined by the spin vector in it's rest frame. In order to use this property we should write the geodesic equation in a locally inertial frame. This can be done with the help of vierbein vectors $e^{(a)}_{\mu}$, where $(a)$ is the Minkowski index. Thus we have the following properties of the vierbein vector
\begin{equation}\label{Neu:Sec1:02}
g_{\mu \nu}=e^{(a)}_{\mu}e^{(b)}_{\nu}\eta _{(a)(b)};\qquad 
\delta ^{\mu}_{\nu}=e^{\mu}_{(a)}e^{(a)}_{\nu};\qquad
\delta ^{(a)}_{(b)}=e^{(a)}_{\mu}e^{\mu}_{(b)};\qquad
\eta _{(a)(b)}=e^{\mu}_{(a)}e^{\nu}_{(b)}g_{\mu \nu}
\end{equation}
where we have the following relation: $e^{\mu}_{(a)}=\eta _{(a)(b)}g^{\mu \nu}e^{(b)}_{\nu}$ for the inverse vierbein and $\eta _{(a)(b)}=\textrm{diag}\left(-1,1,1,1 \right)$ is the metric tensor in the local Minkowski frame. Thus with the use of vierbein vectors the components of spin and four-velocity in a local inertial frame takes the following form, along with the evolution equation
\begin{align}
s^{(a)}&=e^{(a)}_{\mu}S^{\mu};\qquad \frac{ds^{(a)}}{dt}=\frac{1}{\gamma}\mathcal{G}^{(a)(b)}s_{(b)}
\label{Neu:Sec1:03a}
\\
u^{(a)}&=e^{(a)}_{\mu}U^{\mu};\qquad \frac{du^{(a)}}{dt}=\frac{1}{\gamma}\mathcal{G}^{(a)(b)}u_{(b)}
\label{Neu:Sec1:03b}
\end{align}
where, $\gamma =U^{0}=(dt/d\tau)$ and $\mathcal{G}^{(a)(b)}=\eta ^{(a)(c)}\eta ^{(b)(d)}\gamma _{(c)(d)(e)}u^{(e)}=-\mathcal{G}^{(b)(a)}$ is like the electromagnetic field tensor and the object, $\gamma _{(a)(b)(c)}=\eta _{(a)(d)}e_{(b)}^{\mu}e_{(c)}^{\nu}\nabla _{\nu}e^{(d)}_{\mu}$ represents the Ricci rotation coefficients. In order to make the spin evolution in the particle's rest frame, we should apply boost within a locally inertial frame. For that we can introduce two four-vectors $E_{(i)}$ and $B_{(i)}$ constructed from $\mathcal{G}^{(a)(b)}$ such that, $E^{(a)}=\mathcal{G}^{(a)(b)}u_{(b)}$ and $B^{(a)}=(1/2)\epsilon ^{(a)(b)(c)(d)}\mathcal{G}_{(b)(c)}u_{(d)}$. Then both ``electric" field $E_{(a)}$ and ``magnetic" field $B_{(a)}$ has only spatial components denoted by $\mathbf{E}$ and $\mathbf{B}$ such that, $E_{(i)}=\mathcal{G}_{(0)(i)}$ and $G_{(i)(j)}=\epsilon _{(i)(j)(k)}B_{(k)}$. These two vectors will govern the spin evolution in the rest frame of the particle, which is a linear equation in the 
spin 
vector. This spin evolution is similar in spirit to the evolution of a charged particle interacting with external electromagnetic field.

Now we can consider three-dimensional spin vector in the rest frame of the particle given by $\boldsymbol{\zeta}$. Evolution of this three dimensional vector is determined by the following equation:
\begin{align}
\dfrac{d\boldsymbol{\zeta}}{dt}=\frac{2}{\gamma}\left(\boldsymbol{\zeta}\times \mathbf{G}\right)
\end{align}
where the vector $\mathbf{G}$ is constructed out of two three-vectors $\mathbf{E}$ and $\mathbf{B}$ as:
\begin{align}
\mathbf{G}=\frac{1}{2}\left(\mathbf{B}+\frac{1}{1+u^{0}}\left\lbrace\mathbf{E}\times \mathbf{u} \right\rbrace\right)
\end{align}
Then the neutrino spin precession is being determined by the vector $\boldsymbol{\Omega}=\mathbf{G}/\gamma$ and is intimately connected to the spin oscillation frequency of the neutrino.  

Hence the main framework to determine the spin precession vector $\boldsymbol{\Omega}$ goes as follows: (a) Given a spacetime, we need to evaluate the vierbein vectors and the four velocity of the neutrino traveling in this spacetime, (b) Then we need to calculate the covariant derivative in order to determine the tensor $\mathcal{G}^{(a)(b)}$, (c) Starting from the tensor $\mathcal{G}^{(a)(b)}$ calculate the ``electric" and ``magnetic" field and finally (d) Evaluate the three-vector $\mathbf{G}$ in the vierbein frame to get the oscillation frequency. The above calculation has been performed in Appendix \ref{Neu:AppA:A} and we will apply these results in the next section to obtain spin oscillation frequency for the most general static spherically symmetric spacetime.
\section{Neutrino Spin Oscillation in a Static Spacetime With Spherical Symmetry}\label{Neu:Sec:SpinOsc}

The basic equations governing the spin of a particle in a gravitational field has been discussed in \sect{Neu:Sec:SpinIntro}. In this section, we will apply those equations to determine the spin evolution and oscillation of neutrino in a spherically symmetric, static spacetime. This will enable us to determine properties connected to neutrino spin as it propagates around a black hole or a massive object. This question was addressed in the context of general relativity, however in this work we will try to observe the effect of additional correction terms to the \EH action on the spin oscillation of neutrino and possible shift of oscillation frequency. In this section we will outline the basic equations necessary for determination of spin oscillation and then shall concentrate on neutrino flavour oscillation in various alternative theories of gravity before again returning to neutrino spin oscillation.

With possible applications in mind the static spherically symmetric metric ansatz has been chosen in the form:
\begin{equation}\label{Neu:Sec2:01}
ds^{2}=-f(r)dt^{2}+\frac{dr^{2}}{g(r)}+r^{2}d\Omega _{2} ^{2}
\end{equation}
The horizons in the spacetime are determined by the two conditions: $f(r)=0$ and $g(r)=0$. Except for some simple configuration where $f=g$, the two conditions in general do not match. If we choose our coordinates such that, one of them, namely $r$, remains constant on the horizon, then the normal, $r_{a}\propto \nabla _{a}r$, has norm $r_{a}r^{a}=g(r)$. Thus the condition $g(r)=0$ identifies the horizon as a null surface. Similarly the condition $f(r)=0$, identifies the respective surface as a surface of infinite redshift and is known as the ergo surface. For spacetime admitting Killing vector $\xi ^{a}$, for time translation symmetry, the norm of the Killing vector $\xi ^{2}=0$ implies that $f(r)=0$ determines the Killing horizon. Hence both these surfaces have physical significance. 

The oscillation frequency can be obtained in the following steps: First, we need to determine the 4-velocity of the neutrino, in terms of the conserved energy and angular momentum and the vierbein vectors. Then the covariant derivatives of the vierbein vectors are related directly to the ``electric" and ``magnetic" fields. Thus by evaluating the covariant derivatives we have obtained the components of ``electric" and ``magnetic" fields. Then we can construct the vector $\textbf{G}$ and thus the spin precession angular velocity vector $\mathbf{\Omega}$. As we are in a spacetime which is spherically symmetric, we can exploit this symmetry to confine our discussion to the equatorial i.e. $\theta =(\pi /2)$ plane. There only a single component of spin precession survives. This component has distinct expression in different situations. Below we discuss possible situations with respective expressions for the spin precession $\Omega _{2}$ [for detailed calculation see Appendix \ref{Neu:AppA}]:
\begin{itemize}

\item We start our analysis with a situation, where the neutrino is assumed to be massive and is moving in a circular orbit of radius $r_{0}$ around the black hole. Any orbit around a massive object in spherically symmetric spacetime has two constants of motion, namely, the energy $\mathcal{E}$ and the angular momentum $\mathcal{L}$. However for a massive particle it is convenient to work with energy and angular momentum per unit mass, i.e., $\mathcal{E}/m$ and $\mathcal{L}/m$. In the discussion below by energy and angular momentum for a massive particle, will always imply energy and angular momentum per unit mass. 

Given the radius $r_{0}$ of the circular orbit we can determine explicitly the energy per unit mass $E$ and angular momentum per unit mass $L$ in terms of $r_{0}$. The expression for four velocity, energy and angular momentum so obtained is given explicitly in Appendix \ref{Neu:AppB}. Then we can determine the spin precession frequency explicitly with the following expression:
\begin{equation}\label{Neu:Sec2:02}
\Omega _{2}^{\rm{circ}}=-\frac{1}{2}\sqrt{\frac{gf'}{2r_{0}}}\left[1-\frac{r_{0}f'}{\sqrt{2f}
\left(\sqrt{2f}+\sqrt{2f-r_{0}f'}\right)} \right]
\end{equation}
Note that even though we have two functions $f$ and $g$, the spin precession is mostly dominated from the contributions of $f$, while $g$ appears for a ride. Note that in the limit of $g=f=\sqrt{1-(2M/r)}$, we obtain:
\begin{equation}\label{Neu:Sec2:03}
\Omega _{2}^{\rm{circ}}=-\frac{1}{2}\sqrt{\frac{M}{r_{0}^{3}}}\sqrt{1-\frac{3M}{r_{0}}}
\end{equation}
which was derived earlier \cite{Dvornikov2006} for circular orbit in the Schwarzschild spacetime and here verified from our general analysis. Thus we have derived the expression for spin precession frequency for a massive neutrino moving in a circular orbit in the spherically symmetric spacetime of \eq{Neu:Sec2:01}.

\item In the literature quite frequently neutrino is taken to move along null trajectory. Thus it is important to ask, what happens to the spin precession as the neutrino is taken to move on photon circular orbits. This can be obtained directly from \eq{Neu:Sec2:02} by taking appropriate null limit. For circular orbit the null limit is approached when one approaches the photon circular orbit, i.e., circular null geodesics located at $r_{0}=(2f/f')$ (see \eq{Neu:AppB:18} in Appendix \ref{Neu:AppB}). Note that since energy and angular momentum are defined as respective quantities per unit mass the null limit works out finely. It turns out that by taking appropriate limit i.e., $r_{0}\rightarrow (2f/f')$, the spin oscillation frequency identically vanishes, leading to,
\begin{equation}\label{Neu:Sec2:04}
\Omega _{2~(\rm{null})}^{\rm{circ}}=\lim _{r_{0}\rightarrow (2f/f')}-\frac{1}{2}\sqrt{\frac{gf'}{2r_{0}}}\frac{f'}{2f}\left[\frac{2f}{f'}-\frac{r_{0}}{
\left(1+\sqrt{1-r_{0}(f'/2f)}\right)} \right]=0
\end{equation}
The same result can also be argued from the expression for $\Omega _{2}$ obtained explicitly for null trajectory. Following Appendix \ref{Neu:AppB}, it turns out that $\Omega _{2}$ scales as $1/E$, where $E$ is the energy of the neutrino on the null trajectory, which for high energy limit, would tend to the above null result. Note that this result is true for arbitrary static and stationary spacetime. For a very wide class of solutions this result remains true, as we have not used any specific form of the metric keeping the spacetime completely general. 

\item In general, if we consider an orbit with non-zero energy and angular momentum then we have an expression for spin oscillation frequency in general. It turns out that for zero angular momentum the oscillation frequency vanishes. Hence for radial motion there is no oscillation in the spin space and thus the spin of the neutrino remains fixed for such trajectories. Hence for a general timelike orbit the oscillation frequency takes the following form,
\begin{equation}\label{Neu:Sec2:05}
\Omega _{2}^{\rm{geod}}=-\frac{Lf\sqrt{g}}{2Er^{2}}\left[1-\frac{f'Er}{2f\left(E+\sqrt{f}\right)} \right]
\end{equation}  
In the above expression $E$ stands for the energy per unit mass and $L$ accounts for the angular momentum per unit mass of the particle moving on the geodesic. Since the mass of neutrino is very small, it is highly relativistic. Thus we are really interested in the high energy limit of the above expression, which can be obtained by assuming $E\gg L$ and then neglecting higher order terms of the ratio $L/E$. This finally leads to the following expression for spin oscillation frequency:
\begin{equation}\label{Neu:Sec2:06}
\Omega _{2}^{\rm{geod}}\approx -\frac{Lf\sqrt{g}}{2Er^{2}}\left[1-\frac{f'r}{2f} \right]
\end{equation}
From the above expression it is evident that for $r=2f/f'$, $\Omega _{2}^{\rm geod}$ vanishes in the high energy limit. From \eq{Neu:AppB:18} we find that this is precisely the photon circular orbit. Hence our previous result can directly be verified from this line of arguments as well.
\end{itemize}
We have derived the spin oscillation frequency of a neutrino along various trajectories. Among them we have shown that our general result for spin oscillation frequency in circular orbits matches with previous results in the limit of Schwarzschild spacetime. Also we have demonstrated that the oscillation frequency vanishes along null circular trajectories. This was shown earlier in the context of Schwarzschild spacetime, while in this case we have shown the applicability of the result from a more general standpoint. In later sections we shall apply these results to determine helicity flip of neutrino and associated parameter space for its detection.   

\section{Neutrino Flavour Oscillation in Some Classes of Alternative Gravity Theories}\label{Neu:Sec:FlavOsc}

In this section, we will start by first reviewing some basic results about two flavour neutrino oscillation in flat spacetime. The eigenstates of the neutrino in flavour basis is denoted by $\vert \nu _{\textrm{flavour}}\rangle$, which is considered as a superposition of eigenstates in mass basis, denoted by $\vert \nu _{k_{m}}\rangle$. Then the transformation from flavour basis to mass basis is considered to be performed by an unitary transformation $U$ and an associated phase $\exp(-i\Phi _{k_{m}})$, such that, $\Phi _{k_{m}}=E_{k_{m}}t-\overrightarrow{p}_{k_{m}}.\overrightarrow{x}$. Here, $E_{k_{m}}$ and $\overrightarrow{p}_{k_{m}}$ represents the energy and momentum of the $k$-th mass eigenstate. Then for motion of a neutrino from a point A to another point B the phase in general can be interpreted as,
\begin{align}\label{Neu:Sec3:01}
\Phi _{k_{m}}=-\int _{A}^{B}p_{\mu}^{(k_{m})}dx^{\mu}
\end{align}
where, $k_{m}$ signifies the $k$-th mass eigenstate with $m_{k}$ being the corresponding mass. This is considered as generalization of neutrino phase to curved spacetime if we interpret the four-momentum of the neutrino as, $p _{\mu}^{(k_{m})}=m_{k}g_{\mu \nu}(dx^{\nu}/ds)$, where, $g_{\mu \nu}$ is the spacetime metric and $s$ is the proper time along the trajectory of the particle. Then it turns out that in the high energy limit i.e. $E_{k}\gg m_{k}$ along with weak field approximation we arrive at the standard oscillation phase for two flavour neutrino oscillation \cite{Fornengo1997}.

Let us start our analysis by first considering neutrino flavour oscillation in a general static spherically symmetric spacetime. Then after deriving the expressions for oscillation probability and oscillation length in the general case, we will apply these results for some specific situations in alternative theories, which in turn will constrain various parameters of these models. For these purposes we will follow Ref. \cite{Chakraborty2014CQG}. The detailed expressions are being provided in Appendix \ref{Neu:AppA}, we will summarize the important results and shall provide the physical insights behind these results. 

For the general metric ansatz presented in \eq{Neu:AppA:01} we have two conserved quantities the energy and angular momentum. Also the motion is taken to be confined in the equatorial plane. Thus the phase along null geodesic turns out to be:
\begin{equation}\label{Neu:Sec3:02}
\Phi _{k_{m}}^{\rm{null}}=\int ^{B}_{A}dr \left[\frac{m_{k}}{2E_{k_{m}}\sqrt{V}}\sqrt{\frac{f}{g}} \right]
\end{equation}
where, the potential $V$ turns out to have the expression: $V=1-(fL_{k}^{2}/r^{2}E_{k}^{2})$ and the index k merely specifies that we are considering kth massive neutrino. The standard result for phase can be obtained from the high energy limit, with $V\sim 1$ and $f=g=(1-2M/r)$. This leads to the following expression for the null phase:
\begin{equation}\label{Neu:Sec3:03}
\Phi _{k_{m}}^{\rm{null}}=\int _{A}^{B}\frac{m_{k}dr}{2E_{k_{m}}}=\frac{m_{k}^{2}}{2p_{t}^{(k_{m})}}\left(r_{B}-r_{A}\right)
\end{equation}
which exactly matches with the respective result for Schwarzschild spacetime \cite{Fornengo1997}. As the neutrino travels with speed very close to that of light it is considered to have traveled along a null line. However it would be more appropriate to calculate the phase along a timelike trajectory. These two phases differ necessarily as we have taken neutrino to be massive but moving along null geodesic. Then for timelike geodesic the phase turns out to be:
\begin{equation}\label{Neu:Sec3:04}
\Phi _{k_{m}}^{\rm{geod}}=\int _{A}^{B} dr \left[\frac{m_{k}}{\sqrt{E_{k_{m}}^{2}V-f}}\sqrt{\frac{f}{g}}\right]
\approx \int _{A}^{B}\frac{m_{k}dr}{E_{k_{m}}}=2\Phi _{k_{m}}^{\rm{null}}
\end{equation} 
Thus we found that phase along null and timelike trajectory differ by a factor of 2 even for the most general spherically symmetric spacetime discussed here. This factor exists in the flat \cite{Lipkin2000}, Schwarzschild \cite{Bhattacharya1999,Zhang2001}, Kerr-Newmann \cite{Ren2010} and in various alternative gravity theories with solutions having $g_{tt}=g^{rr}$ \cite{Chakraborty2014CQG}. If we neglect the coherence effects, i.e., assume that neutrinos of different flavours are created at the same spacetime point and are also detected at the same spacetime point, then this phase difference is irrelevant for interference. In this particular situation as well we have approximated the trajectory of ultra-relativistic neutrinos as null geodesics which starts and ends at the same spacetime point and thus this factor would not show up in interference. However in principle one could have chosen more complicated trajectories as well (see for example \cite{crocker2004}). 

However as argued in \cite{Giunti2003} this factor appears in flat spacetime due to wrong use of the group velocity in phase. In curved spacetime as well its origin is embedded in the following reason---we are taking the neutrino to be massive, i.e., $p_{a}p^{a}=-m^{2}_{k}$ but at the same time we are approximating it to be moving on a null geodesic since it is ultra relativistic. Due to this \emph{erroneous} use of two separate formulations in a single problem we are getting this factor of 2. 

Having obtained the oscillation phase it is important to consider some non-trivial effect that these alternative theories have on top of the standard Schwarzschild coordinates. The best physical observable for this purpose as pointed out in \cite{Chakraborty2014CQG} is the oscillation length. The oscillation length can be obtained as \cite{Landau1987}
\begin{align}\label{Neu:Sec3:05}
dl=\sqrt{\left(\frac{g_{0a}g_{0b}}{g_{00}}-g_{ab}\right)dx^{a}dx^{b}}
=\frac{dr}{\sqrt{g(r)V}}
\end{align}
Then the differential form of the oscillation phase can be written in terms of differential of proper length from \eq{Neu:Sec3:02} as
\begin{align}\label{Neu:Sec3:06}
d\Phi _{k_{m}}^{\rm{null}}&=\frac{m_{k}dr}{2E_{k_{m}}\sqrt{V}}\sqrt{\frac{f}{g}}
\nonumber
\\
&=\frac{m_{k}\sqrt{f}}{2E_{k_{m}}}dl=\frac{m_{k}^{2}}{2p_{0}^{(k_{m})}}\sqrt{f}dl
\end{align}
The remarkable fact about this equation is that the differential phase does not depend on the $g_{rr}$ component. Thus only the $g_{tt}$ component appears in the oscillation phase and thus affect our solutions. There is another assumption that has gone into the above equation, which corresponds to taking the eigenstates of mass and energy identical. This has been scrutinized critically by several authors and in general it is assumed that $p_{0}^{(k_{m})}$ represents common energy of mass eigenstates. Also there is one critical point that we should stress, as the flat spacetime has both time and space invariance, both $p_{0}$ and $p_{r}$ carry equal momenta. However in curved spacetime $p_{0}$ is a conserved quantity due to existence of timelike Killing vector field, which obviously is not true for $p_{r}$. Hence the equal momentum assumption does not work in curved spacetime \cite{Zhang2001,Grossman1997,Leo2000}. 

Hence the phase shift determining the oscillation turns out to be, $d\Phi _{k}-d\Phi _{j}\propto \Delta m_{kj}^{2}~dl$. Then we can define the oscillation length $L_{\rm{osc}}$ for neutrino in curved spacetime as,
\begin{align}\label{Neu:Sec3:07}
L^{\rm{grav}}_{\rm{osc}}=\frac{dl}{d\left(\frac{\Phi_{kj}^{\rm{null}}}{2\pi} \right)}=\frac{4\pi p_{0}}{\Delta m_{kj}^{2}}\frac{1}{\sqrt{f(r)}}
\end{align}
which for a flat spacetime reduces to, $L_{\rm{osc}}^{\rm{flat}}=(4\pi p_{0}/\Delta m_{kj}^{2})$. Hence finally the fractional change in oscillation length in presence of gravity turns out to be
\begin{align}\label{Neu:Sec3:08}
\Delta l_{1}=\frac{1}{\sqrt{f(r)}}-1
\end{align}
while the other object of interest turns out to be the change in oscillation length in alternative theories in comparison to the corresponding solution in General Relativity. This one has the following expression
\begin{align}\label{Neu:Sec3:09}
\Delta l_{2}=\frac{1}{\sqrt{f_{\rm{alt}}(r)}}-\frac{1}{\sqrt{f_{\rm{GR}}(r)}}
\end{align}
In the above expression $f_{\rm{alt}}(r)$ is the $g_{tt}$ component in the alternative theory and $f_{\rm{GR}}(r)$ is the corresponding $g_{tt}$ element in spacetime described by General Relativity. Next we will try to compute these changes to put bounds on parameters in the theory, which we will compare with the respective parameters using spin oscillation as well. 

\subsection{Dilaton Induced Gravity Theory}

In General Relativity, Schwarzschild solution represents the vacuum solution outside a spherically symmetric massive object. If the mass of the central object becomes comparable to Planck scale then also the same solution describes quite well the structure of the spacetime, except for regions near the massive object (if it forms a black hole, then near the singularity). However when we bring coupling with Maxwell field then the corresponding classical solution, which is the \RN solution gets modified significantly since every solution with non-zero $F_{\mu \nu}$ couples with dilaton. Hence the Maxwell Coupled solution gets drastically modified due to presence of the dilaton field and thus the low energy four dimensional effective action takes the following form as obtained from string theory \cite{Garfinkle1991}:
\begin{equation}\label{Neu:Sec3:10}
\mathcal{A}=\int d^{4}x\sqrt{-g}\left[-R+e^{-2\Phi}F_{\mu \nu}F^{\mu \nu}+2\left(\nabla \Phi \right)^{2} \right] 
\end{equation}
In the above expression, $F_{\mu \nu}=\partial _{\mu}A_{\nu}-\partial _{\nu}A_{\mu}$, is the Maxwell field tensor, belonging to the $U(1)$ subgroup of $E_{8}\times E_{8}$. All other gauge fields except $\Phi$ have been set to zero, then for a purely magnetic Maxwell field we have, the only non-zero component to be, $F^{\theta \phi}=Q\sin \theta$, leading to, $F^{2}=2Q^{2}/R^{4}$. It turns out that only three components of Ricci tensor are non-zero, with one of them satisfying the relation, $R_{00}=-\nabla ^{2}\Phi$. Then the spherically symmetric solution turns out to be \cite{Garfinkle1991}:
\begin{align}\label{Neu:Sec3:11}
ds^{2}=-\left(1-\frac{2M}{r}\right)dt^{2}+\frac{dr^{2}}{\left(1-\frac{2M}{r}\right)}
+r\left(r-e^{2\Phi _{0}}\frac{Q^{2}}{M}\right)d\Omega ^{2}
\end{align}
where, $d\Omega ^{2}=d\theta ^{2}+\sin ^{2}\theta d\phi ^{2}$ is the surface element on $(\theta ,\phi)$ plane, $\Phi _{0}$ represents the asymptotic value of dilaton and $Q$, the black hole charge. This metric resembles the Schwarzschild solution, however with another horizon being located at $r=Q^{2}e^{2\Phi _{0}}/M$. Both the \RN and dilaton solution describes a black hole for small $Q/M$ ratio, while in other situations (with $Q/M<1$) both describe naked singularity. Also from the string theory viewpoint, the strings do not couple to the metric $g_{\mu \nu}$ but rather to $e^{2\Phi}g_{\mu \nu}$. Then in string $\sigma$ model the effective Lagrangian leads to the following action \cite{Garfinkle1991}:
\begin{align}\label{Neu:Sec3:12}
\mathcal{A}=\int d^{4}x \sqrt{-g}e^{-2\Phi}\left[-R-4\left(\nabla \Phi \right)^{2}+F_{\mu \nu}F^{\mu \nu}\right]
\end{align}
From this modified action we obtain the following solution for the metric in spherically symmetric coordinate as \cite{Garfinkle1991}:
\begin{align}\label{Neu:Sec3:13}
ds^{2}=-\frac{1-(2Me^{\Phi _{0}}/\rho)}{1-(Q^{2}e^{3\Phi _{0}}/M\rho)}d\tau ^{2}
+\frac{d\rho ^{2}}{\left[1-(2Me^{\Phi _{0}}/\rho)\right]\left[1-(Q^{2}e^{3\Phi _{0}}/M\rho) \right]}
+\rho ^{2}d\Omega ^{2}
\end{align}
Note that this line element is exactly in the form of the metric ansatz we started given in \eq{Neu:Sec2:01}. The most important parameter in this theory is the dilaton charge, which can be obtained by integrating $\nabla _{\mu}\Phi$ over a two sphere at spatial infinity as \cite{Garfinkle1991,Chakraborty2014PRD8902},
\begin{align}\label{Neu:Sec3:N1}
D=\frac{1}{4\pi}\int d^{2}\Sigma ^{\mu}\nabla _{\mu}\Phi =-\frac{Q^{2}}{2M}e^{2\Phi _{0}}
\end{align}
Having obtained the line element we will now compute the oscillation probability and thus the oscillation length. The most important object that appears in all the expressions correspond to the potential $V(r)$, which in this particular case turns out to have the expression:
\begin{align}\label{Neu:Sec3:14}
V(r)=1-\frac{l_{k_{m}}^{2}}{r^{2}E_{k_{m}}^{2}}\frac{1-(2Me^{\Phi _{0}}/\rho)}{1-(Q^{2}e^{3\Phi _{0}}/M\rho)}
\end{align} 
Hence the phase along null geodesic as well as that along timelike geodesic can be computed in a straightforward manner by just substituting the above expression for potential and corresponding expressions for metric elements in \eq{Neu:Sec3:02} and \eq{Neu:Sec3:04} respectively. Then the two oscillation length, first one corresponding to difference from flat spacetime have the following expression:
\begin{align}\label{Neu:Sec3:15}
\Delta l_{1}=\sqrt{\frac{1-(Q^{2}e^{3\Phi _{0}}/M\rho)}{1-(2Me^{\Phi _{0}}/\rho)}}-1
\end{align}
This expression has very little dependence on the charge $Q$ since for normal astrophysical systems $Q\ll M$. The other comparative oscillation length corresponds to the difference from the general relativistic counterpart i.e. the \RN solution. This comparative oscillation length turns out to have the following expression:
\begin{align}\label{Neu:Sec3:16}
\Delta l_{2}=\sqrt{\frac{1-(Q^{2}e^{3\Phi _{0}}/M\rho)}{1-(2Me^{\Phi _{0}}/\rho)}}
-\frac{1}{\sqrt{\left(1-\frac{2M}{r}+\frac{Q^{2}}{r^{2}} \right)}}
\end{align}
Finally, we can also compute the oscillation probability for an electron type neutrino to remain an electron type neutrino for various lengths. Then we can compare it with the solar neutrino result in order to constrain various parameters of this theory.
\begin{figure*}
\begin{center}

\includegraphics[height=2in, width=3in]{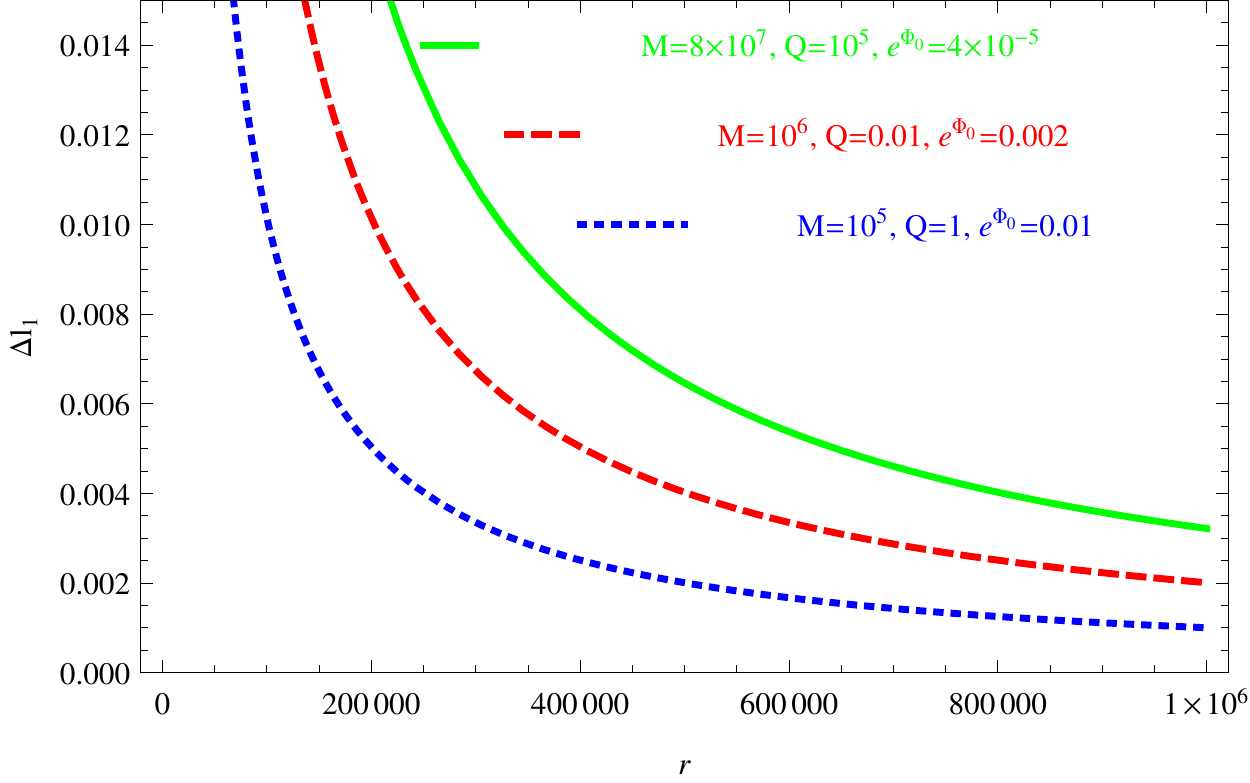}~~
\includegraphics[height=2in, width=3in]{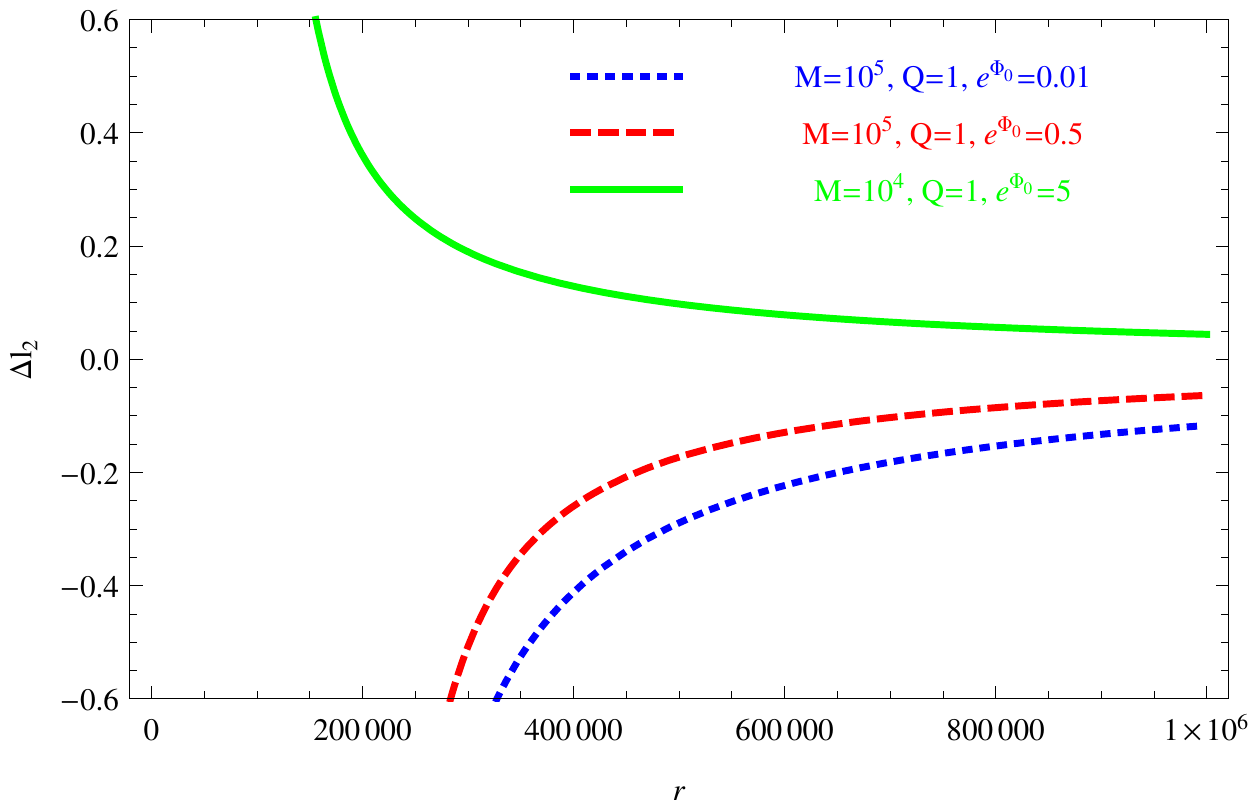}\\

\caption{(color online) In the two figures we have plotted change in neutrino oscillation length induced by dilaton coupled Maxwell Field. The left figure shows variation of $\Delta l_{1}$ difference of neutrino oscillation length from the dilaton coupled theory to that of flat spacetime with radial distance from the source. Thus as we move outwards the difference goes to zero, since at large distance from the source the spacetime becomes flat. While the second figure depicts variation of $\Delta l_{2}$ the difference from Schwarzschild geometry. Also at large distance this difference approaches zero, showing at large scale the solution resembling Schwarzschild spacetime.}\label{Neu:Fig:01}

\end{center}
\end{figure*}

\begin{figure*}
\begin{center}

\includegraphics[height=3in, width=5in]{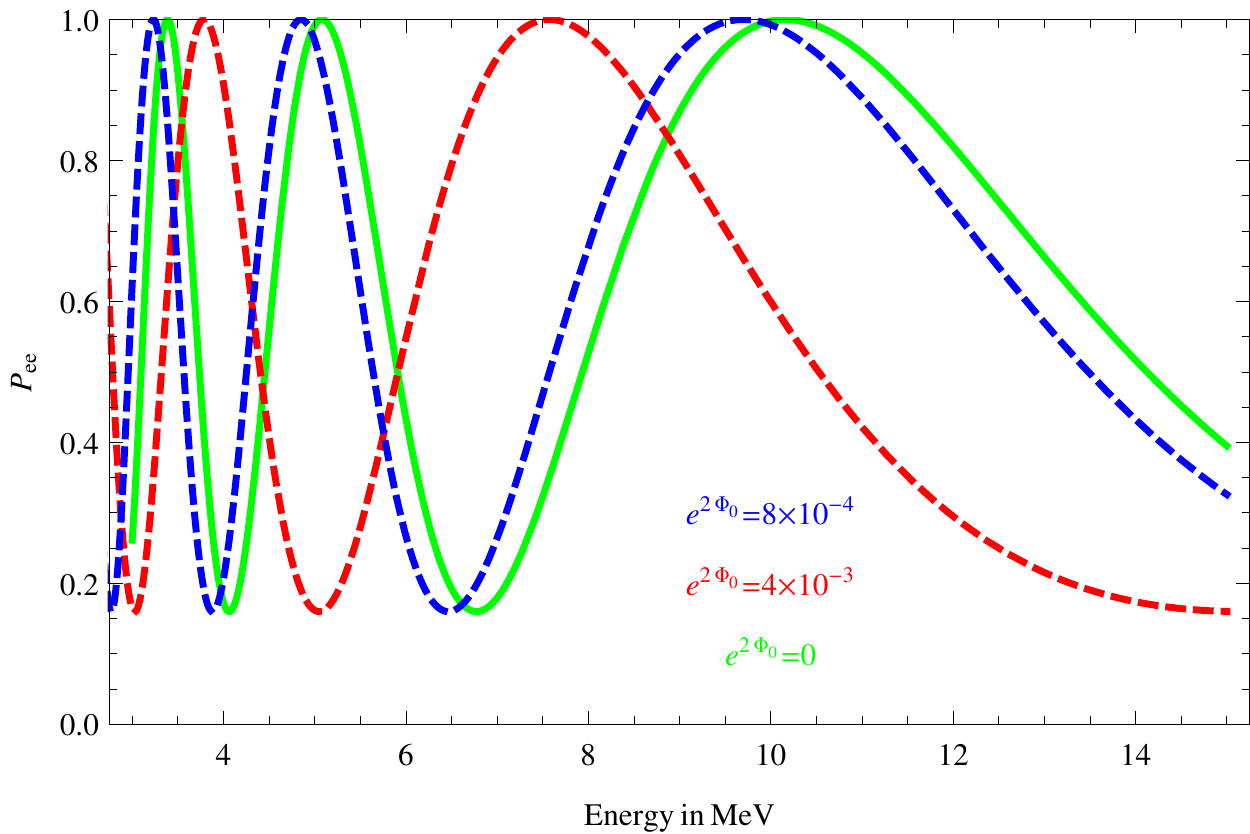}

\caption{(color online) In this figure we have depicted the probability of an electron type neutrino to remain an electron type neutrino with its energy in MeV for a length of 180 km. Different curves describe the transition probability corresponding to different values of $e^{2\Phi _{0}}$, which depends on the asymptotic value of the dilaton. Green curve represents oscillation probability in absence of dilaton field, while the other two curves depict oscillation probability with dilaton field being present. }\label{Neu:Fig:02}

\end{center}
\end{figure*}
The variation of the two comparative oscillation lengths with radial distance have been presented in Fig. \ref{Neu:Fig:01}. Also the probability for an electron type neutrino to remain another electron type neutrino for a definite energy window has been presented in Fig. \ref{Neu:Fig:02}. The fact that we are considering probability of an electron type neutrino to remain an electron type neutrino is illustrated in Fig. \ref{Neu:Fig:02} as the maximum probability reaches the value unity. 

However we can do more, from the neutrino oscillation probability, in this gravity model we can constrain the parameter $D$, defined through \eq{Neu:Sec3:N1} using the data for solar neutrino oscillation. This can be performed along the following lines, since the geometry and hence the oscillation probabilities are different, the theoretical flux would differ from the Schwarzschild value and this difference depends upon the dilaton charge. To be compatible with the experiment the excess flux must be within the statistical errors present in observed fluxes. This in turn will provide stringent constraints on the dilaton charge $D$. The constraints on the dilaton charge for various solar neutrino experiments have been presented in Table \ref{Neu_Osc_DT01}.
\begin{table}
\begin{center}
\caption{\bf Results from real time experiments regarding $^{8}B$ solar neutrino flux have been shown. The errors presented are statistical errors. Bounds on the dilaton charge $D$ from each of these experiments have been estimated.}\label{Neu_Osc_DT01}
\centering

\begin{tabular}{|c|c|c|c|}

\hline
\hline
{\bf Experiment} & {\bf Reaction} & {\bf $^{8}B$ $\nu$ flux} & {\bf Bound on dilaton} \\[0.3ex]

{} & {} & {} & {\bf charge $D$} \\[0.3ex]

\hline
\hline

Kamiokande \cite{Fukuda1996}
&
$\nu e$
&
$2.80 \pm 0.19$
&
$< 2.12 \times 10^{-9}$\\

Super-K I \cite{Hosaka2006}
&
$\nu e$
&
$2.38\pm 0.02$
&
$< 2.05 \times 10^{-9}$\\

Super-K II \cite{Cravens2008}
&
$\nu e$
&
$2.41 \pm 0.05$
&
$< 2.07 \times 10^{-9}$\\

Super-K III \cite{Abe2011}
&
$\nu e$
&
$2.32 \pm 0.04$
&
$< 1.94 \times 10^{-9}$\\

SNO Phase I \cite{Ahmad2002}
&
CC
&
$1.76^{+0.06}_{-0.05}$
&
$< 1.74 \times 10^{-9}$\\

~~~~(pure $D_{2}O$)
&
$\nu e$
&
$2.39^{+0.24}_{-0.23}$
&
$< 2.05 \times 10^{-9}$\\

&
NC
&
$5.09^{+0.44}_{-0.43}$
&
$< 3.76 \times 10^{-9}$\\

SNO Phase II \cite{Aharmin2005}
&
CC
&
$1.68 \pm 0.06$
&
$< 1.69 \times 10^{-9}$\\

~~~~(NaCl in $D_{2}O$)
&
$\nu e$
&
$2.35 \pm 0.22$
&
$< 1.98 \times 10^{-9}$\\

&
NC
&
$4.94 \pm 0.21$
&
$< 3.53 \times 10^{-9}$\\

SNO Phase III \cite{Aharmin2008}
&
CC
&
$1.67^{+0.05}_{-0.04}$
&
$< 1.68 \times 10^{-9}$\\

~~~~~($^{3}He$ counters)
&
$\nu e$
&
$1.77 ^{+0.24}_{-0.21}$
&
$< 1.74 \times 10^{-9}$\\

&
NC
&
$5.54^{+0.33}_{-0.31}$
&
$< 3.92 \times 10^{-9}$\\

Borexino \cite{Bellini2010}
&
$\nu e$
&
$2.4 \pm 0.4$
&
$< 2.07 \times 10^{-9}$\\

\hline
\hline

\end{tabular}
\end{center}
\end{table}

\subsection{Quadratic Gravity Theory}

In high energy physics \gr is often treated as the low energy approximation of some underlying fundamental theory. This idea is reconciled by adding higher curvature terms to \gr and observing their implications in high energy phenomenon through astrophysical experiments. For example, modification of accretion disk structure in presence of higher curvature term has been discussed in Ref. \cite{Chakraborty2015CQG}. In this section we consider a class of alternative theories in four dimensions, which is obtained by modifying the \EH action through introduction of various quadratic and algebraic curvature scalars with proper couplings, such that the action gets modified to \cite{Yunes2011,Chakraborty2014PRD8902,Chakraborty2013}:
\begin{align}\label{Neu:Sec3:17}
\mathcal{A}&=\int d^{4}x \sqrt{-g}\Big[\frac{R}{16\pi G}+\alpha _{1}f_{1}(\upsilon)R^{2}+\alpha _{2}f_{2}(\upsilon)R_{\mu \nu}R^{\mu \nu}+\alpha _{3}f_{3}(\upsilon)R_{\mu \nu \alpha \beta}R^{\mu \nu \alpha \beta}
\nonumber
\\
&+\alpha _{4}f_{4}(\upsilon)R_{\mu \nu \alpha \beta}^{*}R^{\mu \nu \alpha \beta}-\frac{\beta}{2}\left\lbrace \nabla _{a}\upsilon \nabla ^{a}\upsilon +2V(\upsilon)\right\rbrace +L_{\rm{matter}}\Big]
\end{align}
where $g$ represents the determinant of the metric $g_{\mu \nu}$. Among other quantities we have the Ricci scalar R, Ricci tensor $R_{\mu \nu}$, Riemann tensor $R_{\mu \nu \alpha \beta}$ along with its dual $R^{*}_{\mu \nu \alpha \beta}$ constructed from the metric $g_{\alpha \beta}$. In the above action $L_{\rm{matter}}$ represents the matter Lagrangian, $\upsilon$ is an arbitrary scalar field with $\left(\alpha _{i},\beta \right)$ representing coupling constants. These theories have their motivation in the low energy expansion of string theory \cite{Boulware1985,Green1987}. Then by varying the above action field equation with respect to $g_{\mu \nu}$ can be obtained, to which spherically symmetric solution can be arrived at through the following metric ansatz
\begin{align}\label{Neu:Sec3:18}
ds^{2}-f_{0}\left[1+\epsilon h_{0}(r)\right]dt^{2}+f_{0}^{-1}\left[1+\epsilon k_{0}(r)\right]dr^{2}
+r^{2}d\Omega ^{2}
\end{align}
where $\epsilon$ comes from expansion of the scalar field around its solution and $f_{0}=1-(2M_{0}/r)$. Here $M_{0}$ represents the bare mass and $d\Omega ^{2}$ is the two dimensional surface element. Also in the above equation $h_{0}$ and $k_{0}$ represents small deformation around the Schwrazschild value. Then we can use solution of scalar field equation in order to obtain a solution to the modified field equation at linear order in $\epsilon$. Further the requirement that metric should be asymptotically flat and regular at $r=2M_{0}$ can lead to unique solution. Then defining a dimensionless coupling parameter, $\zeta =(16\pi G \alpha _{3}^{2}/\beta M_{0}^{4})$, we can introduce a physical mass parameter given by $M=M_{0}\left[1+(49/80)\zeta \right]$. Then the modified line element takes the following form \cite{Yunes2011,Chakraborty2014PRD8902}:
\begin{align}\label{Neu:Sec3:19}
ds^{2}=-f(r)\left[1+\frac{\zeta}{3f(r)}\left(\frac{M}{r}\right)^{3}h(r)\right]dt^{2}
+\frac{1}{f(r)}\left[1-\frac{\zeta}{f(r)}\left(\frac{M}{r}\right)^{2}k(r)\right]dr^{2}
+r^{2}d\Omega ^{2}
\end{align}
where we have $f(r)=1-(2M/r)$, and the two other unknown functions $h(r)$ and $k(r)$ have the following expressions:
\begin{align}
h(r)&=1+\frac{26M}{r}+\frac{66}{5}\frac{M^{2}}{r^{2}}+\frac{96}{5}\frac{M^{3}}{r^{3}}-\frac{80M^{4}}{r^{4}}
\label{Neu:Sec3:20a}
\\
k(r)&=1+\frac{M}{r}+\frac{52}{3}\frac{M^{2}}{r^{2}}+\frac{2M^{3}}{r^{3}}+ \frac{16M^{4}}{5r^{4}}- \frac{368}{3}\frac{M^{5}}{r^{5}}
\label{Neu:Sec3:20b}
\end{align}
The important point to be stressed is that the metric elements so obtained do not depend on the bare mass $M_{0}$ but on the physical mass $M$. Thus having obtained the solutions we will now consider the oscillation length and oscillation phase of neutrino in this spacetime. 

\begin{figure*}
\begin{center}

\includegraphics[height=2in, width=3in]{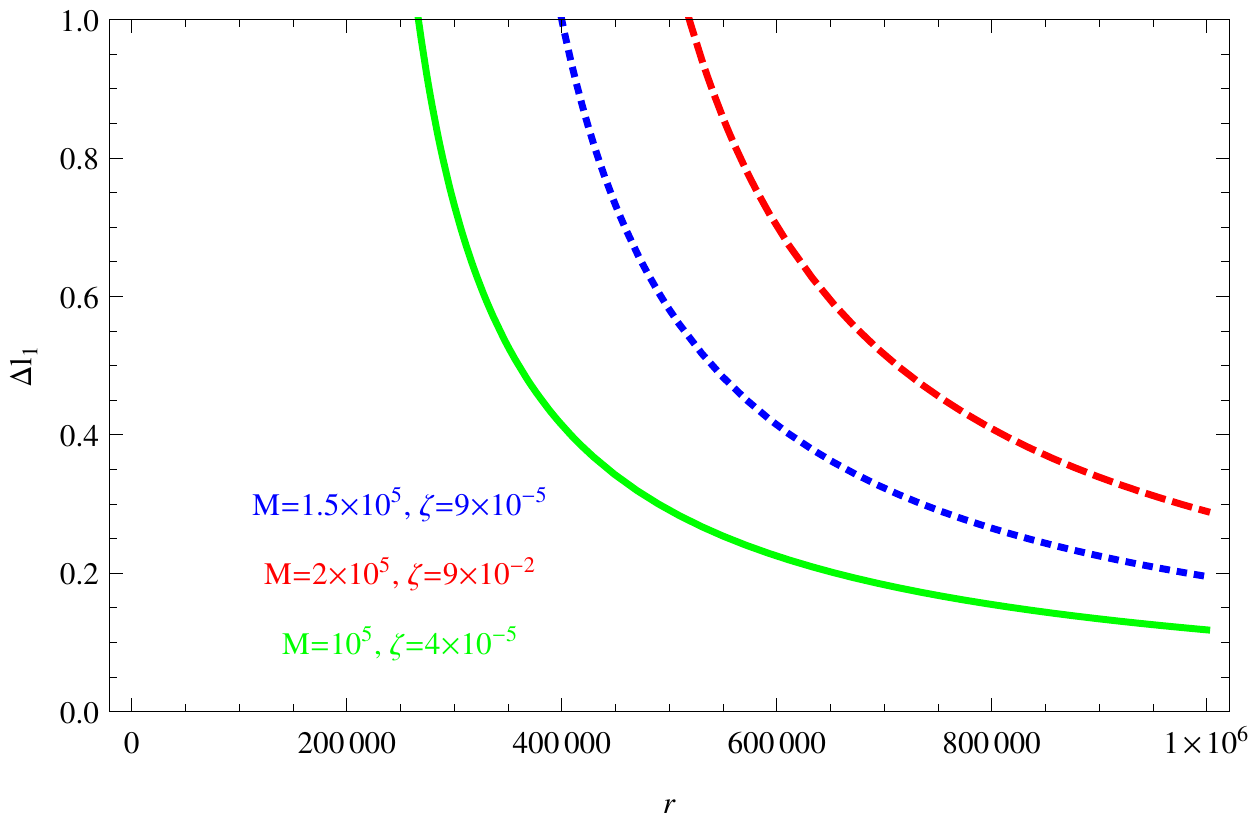}~~
\includegraphics[height=2in, width=3in]{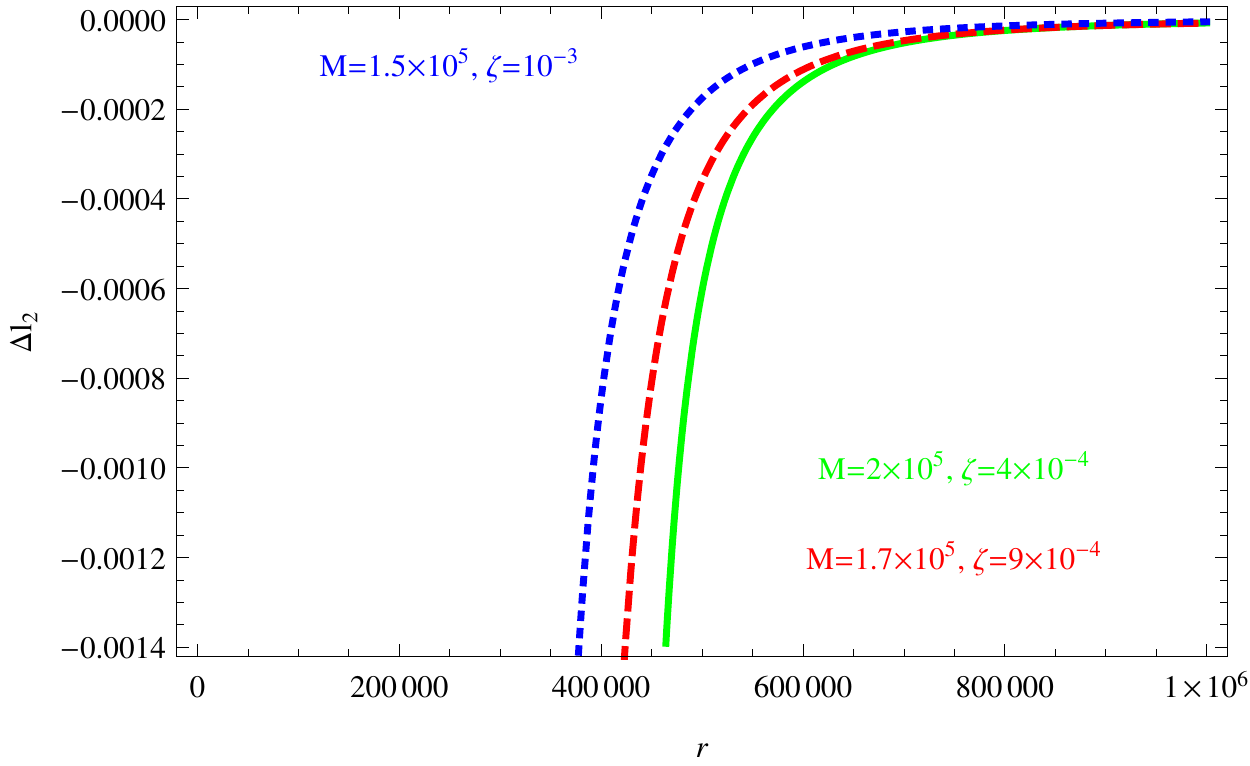}\\

\caption{(color online) In this figure the variation of comparative oscillation length $\Delta l_{1}$ and $\Delta l_{2}$ with radial coordinate have been presented for various choice of parameters. The left figure shows variation of $\Delta l_{1}$, the difference of oscillation length in quadratic gravity and flat spacetime. On the other hand, the second figure depicts variation of $\Delta l_{2}$, difference from Schwarzschild geometry. At large distances this comparative difference tends to zero resembling Schwarzschild behavior.}\label{Neu:Fig:03}

\end{center}
\end{figure*}

\begin{figure*}
\begin{center}

\includegraphics[height=3in, width=5in]{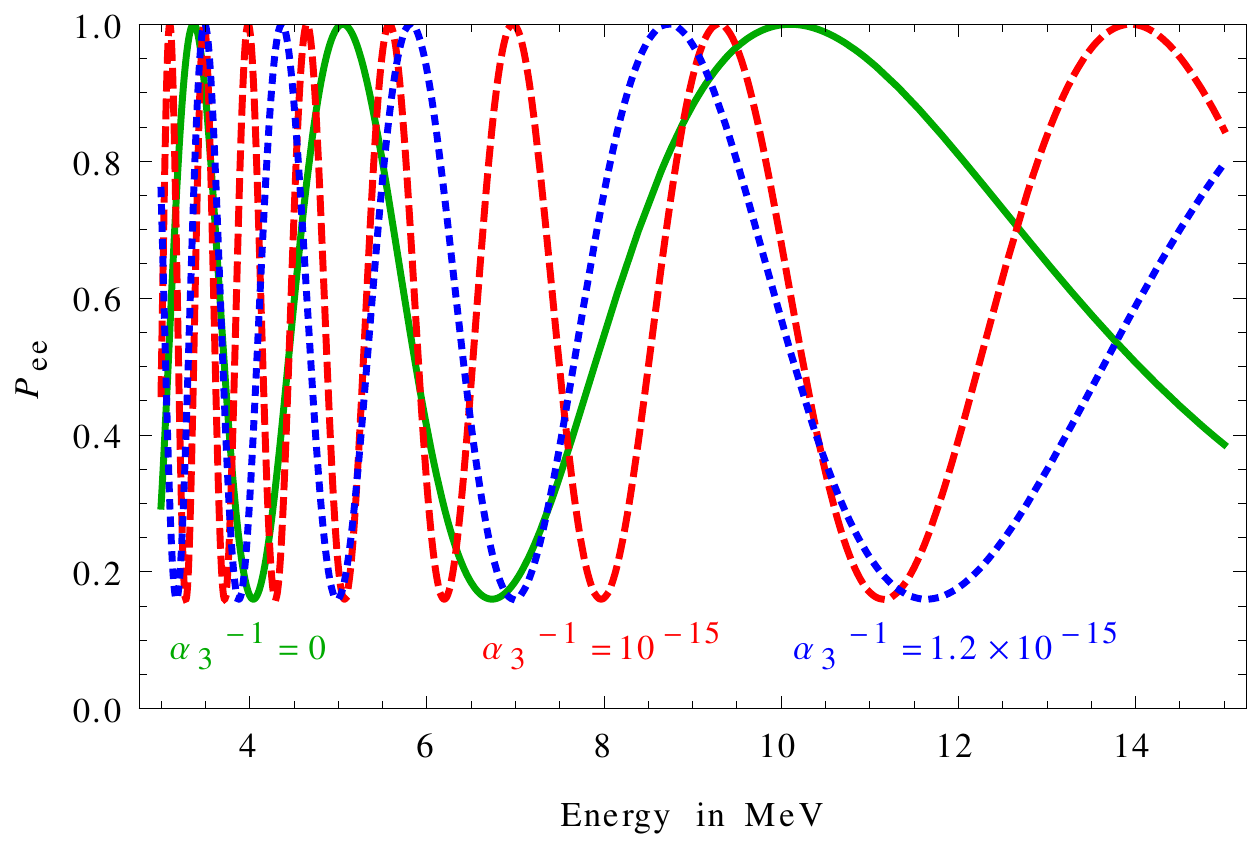}

\caption{(color online) In this figure we have depicted the probability of an electron type neutrino to remain an electron type neutrino with its energy in MeV for a length of 180 km. Different curves describe the probability of this event corresponding to different values of $\alpha _{3}^{-1}$. Green curve represents oscillation probability in absence of quadratic corrections, while the other two curves depict oscillation probability with quadratic corrections being present.}\label{Neu:Fig:04}

\end{center}
\end{figure*}

The line element is exactly in the form of the metric ansatz we started with as given in \eq{Neu:Sec2:01}. Having arrived at the line element we will now compute the oscillation probability and thus the oscillation length. The important expression that appears in all the expressions for oscillation phase is the potential $V(r)$, which in this case leads to the following expression:
\begin{align}\label{Neu:Sec3:21}
V(r)=1-\frac{l_{k_{m}}^{2}}{r^{2}E_{k_{m}}^{2}}f(r)\left[1+\frac{\zeta}{3f(r)}\left(\frac{M}{r}\right)^{3}h(r)\right]
\end{align} 
where, $f(r)=1-(2M/r)$ and $h(r)$ is given by \eq{Neu:Sec3:20a}. Then the phase along null geodesic as well as that along timelike geodesic can be computed directly by substituting the above expression for potential and corresponding expressions for metric elements in \eq{Neu:Sec3:02} and \eq{Neu:Sec3:04} respectively. Then the two oscillation length, first one related to the difference from flat spacetime have the following expression:
\begin{align}\label{Neu:Sec3:22}
\Delta l_{1}=\frac{1}{\sqrt{f(r)\left[1+\frac{\zeta}{3f(r)}\left(\frac{M}{r}\right)^{3}h(r)\right]}}-1
\end{align}
while the other one corresponding to the difference from the general relativistic counterpart i.e. the \RN solution turns out to be
\begin{align}\label{Neu:Sec3:23}
\Delta l_{2}=\frac{1}{\sqrt{f(r)\left[1+\frac{\zeta}{3f(r)}\left(\frac{M}{r}\right)^{3}h(r)\right]}}
-\frac{1}{\sqrt{1-\frac{2M}{r}}}
\end{align}
In both the expressions, $f(r)=1-(2M/r)$ and $h(r)$ is given by \eq{Neu:Sec3:20a}. Finally, we can compute the oscillation probability of electron type neutrino converting to electron type neutrino which we can compare with the solar neutrino result in order to constrain various parameters of this theory. Even though the parameter $\zeta$ appears explicitly in the metric elements, it is not a fundamental parameter. The fundamental parameter is $\alpha _{3}^{-1}$, which is related to $\zeta$ via $\zeta =(16\pi G \alpha _{3}^{2} M_{0}^{4})$. Figure \ref{Neu:Fig:03} depicts the difference lengths $\Delta l_{1}$ and $\Delta l_{2}$ for various choice of parameters, while figure \ref{Neu:Fig:04} depicts oscillation probability for various choices of $\alpha _{3}^{-1}$. In Table \ref{Neu_Osc_QT02} we present constraints on $\alpha _{3}^{-1}$ from solar neutrino experiments using identical techniques as explained in previous section.
\begin{table}
\begin{center}
\caption{\bf Results from real time experiments regarding $^{8}B$ solar neutrino flux have been shown. The errors presented are statistical errors. Bounds on $\alpha _{3}^{-1}$ from each of these experiments have been estimated.}\label{Neu_Osc_QT02}
\centering

\begin{tabular}{|c|c|c|c|}

\hline
\hline
{\bf Experiment} & {\bf Reaction} & {\bf $^{8}B$ $\nu$ flux} & {\bf Bound on } \\[0.3ex]

{} & {} & {} & {\bf $\alpha _{3}^{-1}$} \\[0.3ex]

\hline
\hline

Kamiokande \cite{Fukuda1996}
&
$\nu e$
&
$2.80 \pm 0.19$
&
$< 0.95 \times 10^{-18}$\\

Super-K I \cite{Hosaka2006}
&
$\nu e$
&
$2.38\pm 0.02$
&
$< 0.42 \times 10^{-18}$\\

Super-K II \cite{Cravens2008}
&
$\nu e$
&
$2.41 \pm 0.05$
&
$< 0.45 \times 10^{-18}$\\

Super-K III \cite{Abe2011}
&
$\nu e$
&
$2.32 \pm 0.04$
&
$< 0.36 \times 10^{-18}$\\

SNO Phase I \cite{Ahmad2002}
&
CC
&
$1.76^{+0.06}_{-0.05}$
&
$< 9.16 \times 10^{-19}$\\

~~~~(pure $D_{2}O$)
&
$\nu e$
&
$2.39^{+0.24}_{-0.23}$
&
$< 0.43 \times 10^{-18}$\\

&
NC
&
$5.09^{+0.44}_{-0.43}$
&
$< 2.32 \times 10^{-18}$\\

SNO Phase II \cite{Aharmin2005}
&
CC
&
$1.68 \pm 0.06$
&
$< 8.23 \times 10^{-19}$\\

~~~~(NaCl in $D_{2}O$)
&
$\nu e$
&
$2.35 \pm 0.22$
&
$< 0.39 \times 10^{-18}$\\

&
NC
&
$4.94 \pm 0.21$
&
$< 2.16 \times 10^{-18}$\\

SNO Phase III \cite{Aharmin2008}
&
CC
&
$1.67^{+0.05}_{-0.04}$
&
$< 8.12 \times 10^{-19}$\\

~~~~~($^{3}He$ counters)
&
$\nu e$
&
$1.77 ^{+0.24}_{-0.21}$
&
$< 9.19 \times 10^{-19}$\\

&
NC
&
$5.54^{+0.33}_{-0.31}$
&
$< 2.58 \times 10^{-18}$\\

Borexino \cite{Bellini2010}
&
$\nu e$
&
$2.4 \pm 0.4$
&
$< 0.44 \times 10^{-18}$\\

\hline
\hline

\end{tabular}
\end{center}
\end{table}

\subsection{Regular Black holes}

Existence of singularity appears to be an inherent property of most of the solutions to gravitational field equation in general relativity. This problem is generally avoided by considering cosmic censorship conjecture according to which singularities are always dressed by event horizons. This shows that any pathological behavior at the singularity has no influence on the exterior region. To circumvent these difficulties a set of regular black hole solutions were proposed, known as ``Bardeen black holes" \cite{Borde1997}. However none of these models is an exact solution to Einstein's equation with some known physical source associated. It was first suggested in \cite{ABG1998,ABG1999a,ABG1999b} that even within the context of general relativity it is possible to construct singularity free solutions. However this can be achieved only at the price of introducing non-linear sources. Thus by introducing non-linear electrodynamics to Einstein gravity it is possible to obtain singularity free solutions. The 
solution so obtained has the line element with $f(r)=g(r)$, where the function $f(r)$ has the following expression:
\begin{enumerate}
\item Bardeen spacetime
\begin{equation}
f(r)=1-\frac{2mr^{2}}{\left(q^{2}+r^{2}\right)^{3/2}}
\end{equation}
\item ABG spacetime
\begin{equation}
f(r)=1-\frac{2mr^{2}}{\left(q^{2}+r^{2}\right)^{3/2}}+\frac{q^{2}r^{2}}{\left(q^{2}+r^{2}\right)^{2}}
\end{equation}
\end{enumerate}
In the above expressions $m$ denotes the standard gravitational mass and $q$ stands for charge parameter measured in units of mass $m$. Since both the line elements are exactly in the form presented in \eq{Neu:Sec2:01} we can carry forward our analysis presented in earlier sections. The important expression appearing in all the related expressions is the potential term $V(r)$ having the expression:
\begin{enumerate}
\item Bardeen spacetime
\begin{align}
V(r)=1-\frac{L_{k}^{2}}{r^{2}E_{k}^{2}}\left(1-\frac{2mr^{2}}{\left(q^{2}+r^{2}\right)^{3/2}}\right)
\end{align}
\item ABG spacetime
\begin{align}
V(r)=1-\frac{L_{k}^{2}}{r^{2}E_{k}^{2}}\left[1-\frac{2mr^{2}}{\left(q^{2}+r^{2}\right)^{3/2}}+\frac{q^{2}r^{2}}{\left(q^{2}+r^{2}\right)^{2}}\right]
\end{align}
\end{enumerate}
Then the above expressions when substituted in \eqs{Neu:Sec3:02} and (\ref{Neu:Sec3:04}) the respective expressions for oscillation phase along null and timelike geodesics can be obtained. Also the two oscillation lengths, one related to the oscillation length difference between flat spacetime and the curved spacetime has the following expression:
\begin{enumerate}
\item Bardeen spacetime
\begin{align}
\Delta l_{1}=\frac{1}{\sqrt{1-\frac{2mr^{2}}{\left(q^{2}+r^{2}\right)^{3/2}}}}-1
\end{align}
\item ABG spacetime
\begin{align}
\Delta l_{1}=\frac{1}{\sqrt{1-\frac{2mr^{2}}{\left(q^{2}+r^{2}\right)^{3/2}}+\frac{q^{2}r^{2}}{\left(q^{2}+r^{2}\right)^{2}}}}-1
\end{align}
\end{enumerate}
The second one represents the difference between the oscillation length in the spacetime governed by alternative gravity theories and corresponding \gr\ solution. This has the following expression:
\begin{enumerate}
\item Bardeen spacetime
\begin{align}
\Delta l_{2}=\frac{1}{\sqrt{1-\frac{2mr^{2}}{\left(q^{2}+r^{2}\right)^{3/2}}}}-\frac{1}{\sqrt{1-\frac{2m}{r}+\frac{q^{2}}{r^{2}}}}
\end{align}
\item ABG spacetime
\begin{align}
\Delta l_{2}=\frac{1}{\sqrt{1-\frac{2mr^{2}}{\left(q^{2}+r^{2}\right)^{3/2}}+\frac{q^{2}r^{2}}{\left(q^{2}+r^{2}\right)^{2}}}}-\frac{1}{\sqrt{1-\frac{2m}{r}+\frac{q^{2}}{r^{2}}}}
\end{align}
\end{enumerate}

\begin{figure*}
\begin{center}

\includegraphics[height=2in, width=3in]{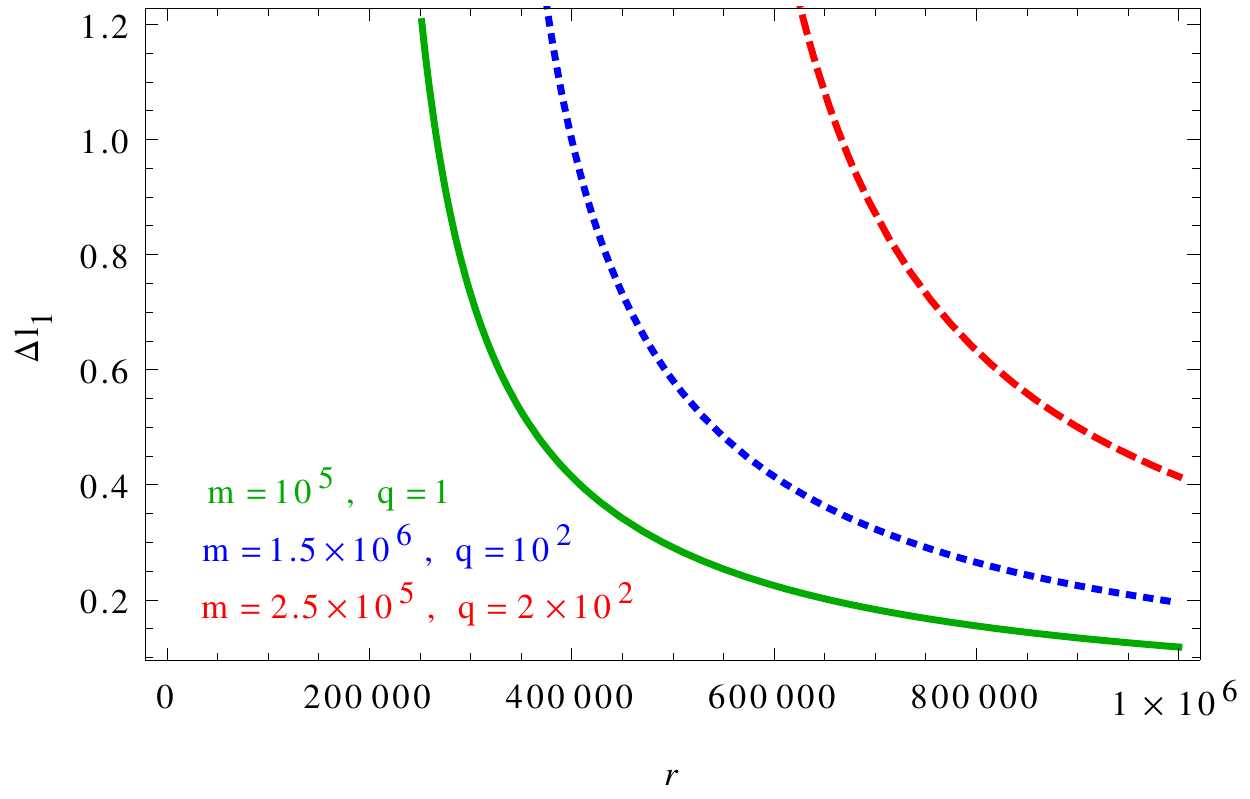}~~
\includegraphics[height=2in, width=3in]{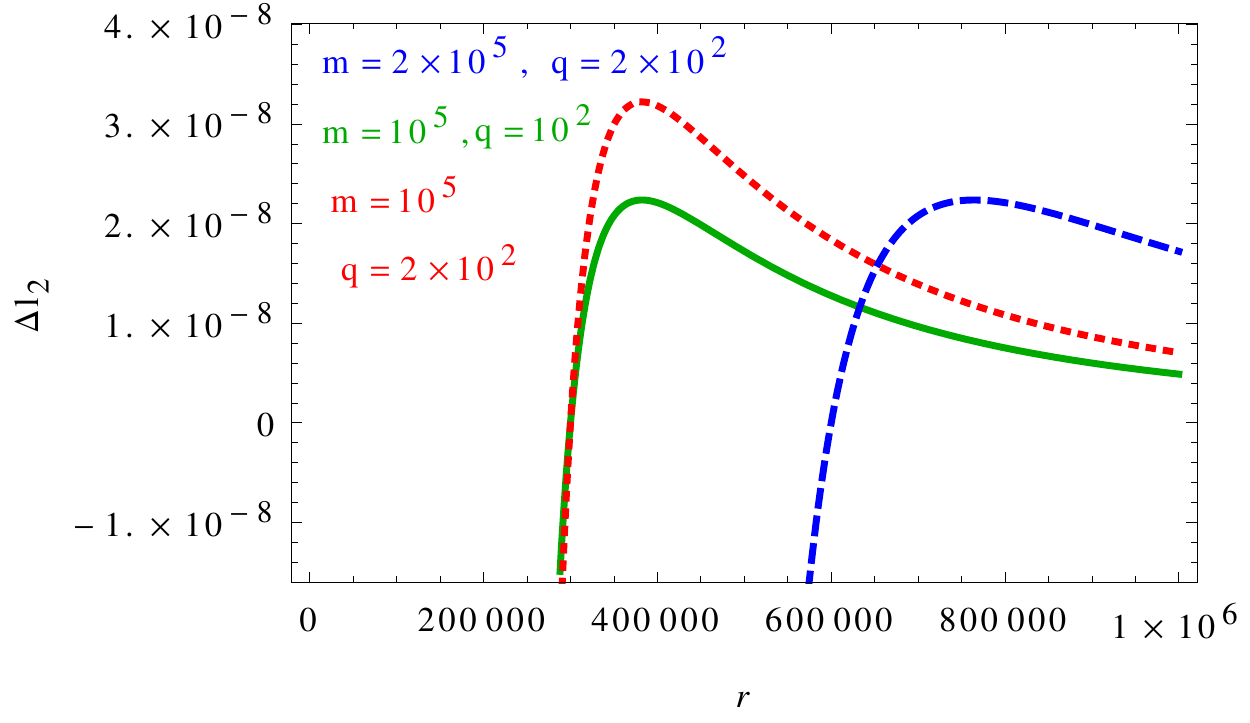}\\

\caption{(color online) In this figure the variation of comparative oscillation length $\Delta l_{1}$ and $\Delta l_{2}$ with radial coordinate has been presented for various choices of parameters. The left figure shows variation of $\Delta l_{1}$, the difference of oscillation length in Bardeen spacetime and flat spacetime. On the other hand the second figure depicts variation of $\Delta l_{2}$, difference from \RN geometry. At large distance this comparative difference tends to zero resembling \RN behavior.}\label{Neu:Fig:05}

\end{center}
\end{figure*}
\begin{figure*}
\begin{center}

\includegraphics[height=1.8in, width=3in]{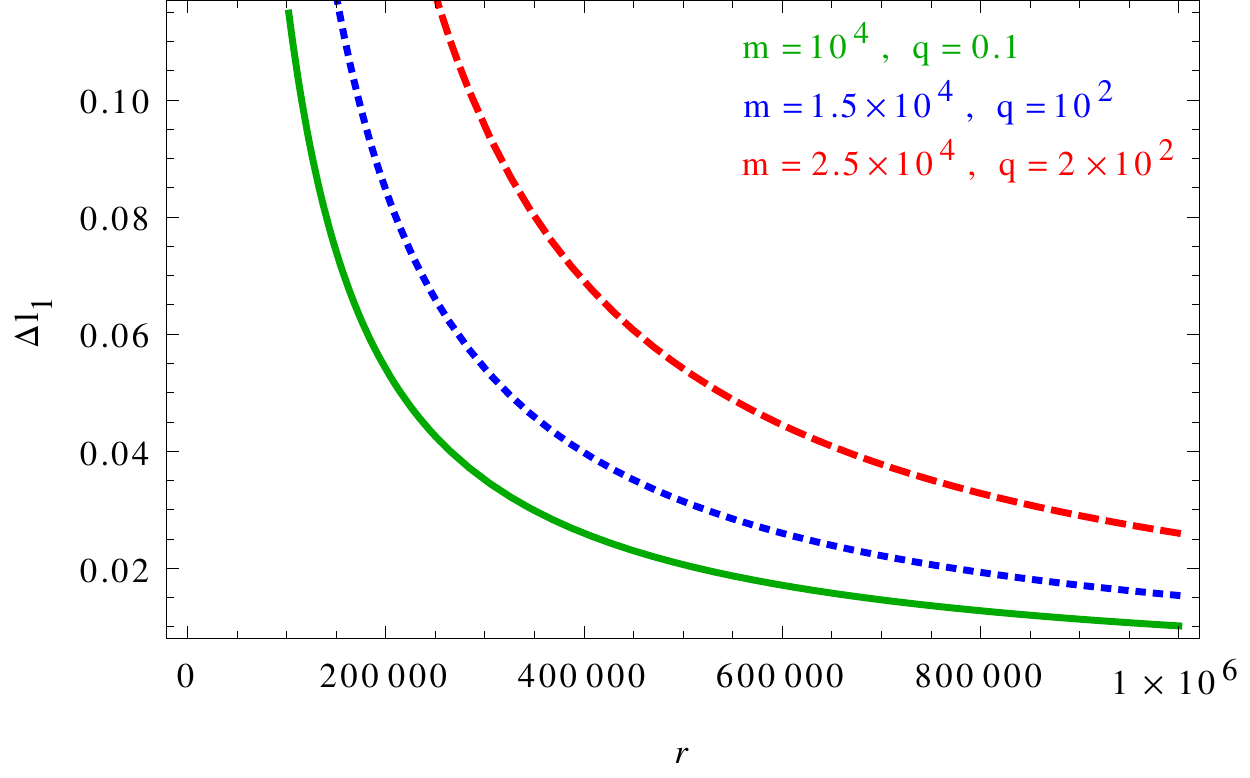}~~
\includegraphics[height=1.8in, width=3in]{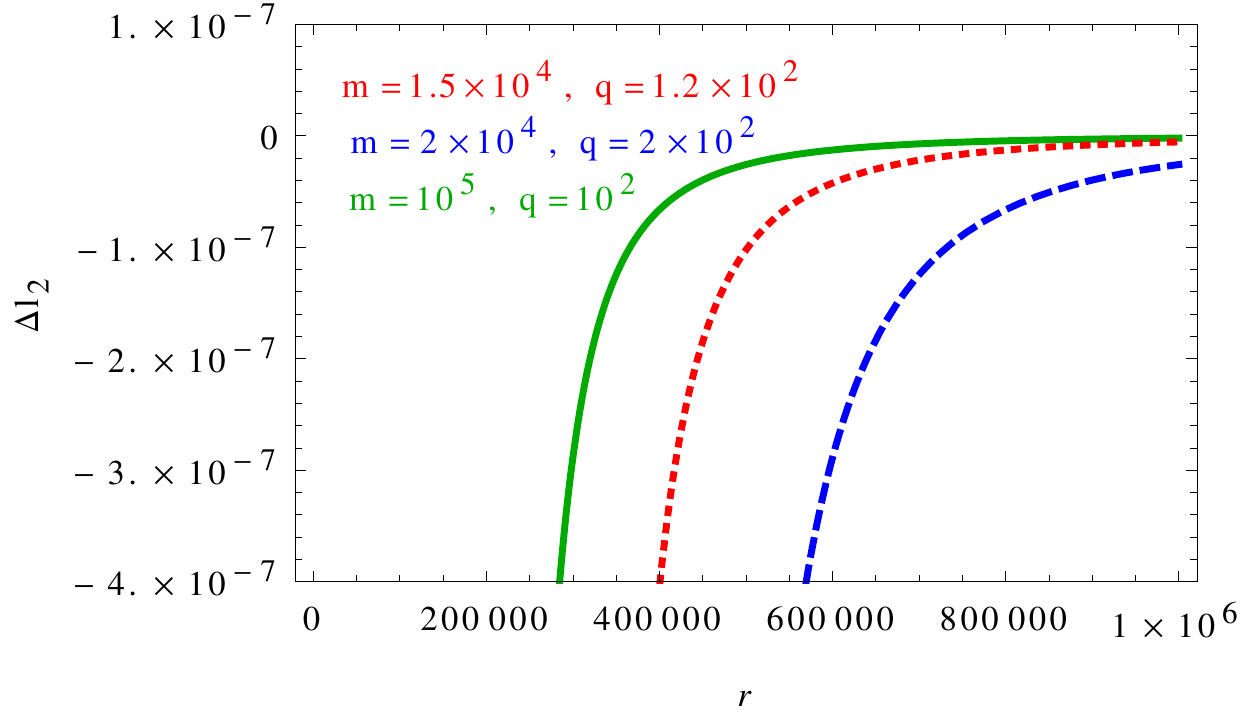}\\

\caption{(color online) In this figure the variation of comparative oscillation length $\Delta l_{1}$ and $\Delta l_{2}$ with radial coordinate has been presented for various choices of parameters in ABG spacetime. The left figure shows variation of $\Delta l_{1}$ and the right figure depicts variation of $\Delta l_{2}$. See text for more discussions.}\label{Neu:Fig:06}

\end{center}
\end{figure*}
Both the comparative behaviors regrading oscillation lengths have been illustrated. Figure \ref{Neu:Fig:05} depicts the comparative oscillation lengths $\Delta l_{1}$ and $\Delta l_{2}$ for Bardeen spacetime and figure \ref{Neu:Fig:06} shows identical diagrams but for ABG spacetime. Having obtained all the oscillation lengths and their comparative behavior, we can now compute the oscillation probability of electron type neutrino converting to electron type neutrino. It turns out that the corrections to the oscillation length and hence the departure of oscillation probability in these regular black hole solutions from that in the Schwarzschild spacetime is quite small. They cannot be used accurately to place tight constraints on the parameter $q$ as we have done in the previous two situations. This originates from the fact that all the corrections are sub-leading and falls off much faster, leaving very little correction in comparison to the previous two cases.

Having discussed three alternative gravity theories and the neutrino flavour oscillation thereof, we will now concentrate on neutrino helicity flip, i.e., spin oscillation of neutrino in alternative theories. 
\section{Neutrino Helicity Flip in Alternative Gravity Theories}\label{Neu:Sec:Hel}

In \sect{Neu:Sec:SpinOsc} we have derived the spin oscillation frequency for neutrino, both along circular and non-circular geodesics. We have observed that for circular motion in the high energy limit, the oscillation frequency vanishes. This is a \emph{very important} result, it suggests that a neutrino moving in a circular orbit does not undergo any change of its spin and hence helicity (since neutrino is always a highly relativistic particle). More importantly this result is true not only in Einstein gravity but in all other alternative gravity theories, as the result essentially depends only on static and spherical symmetry. This result was first derived for Schwarzschild spacetime in \cite{Dvornikov2006} and in this work is shown to \emph{transcend} general relativity. 

To start with we assume that the neutrino initially is left-handed, implying its spin vector being anti-parallel to the velocity of the particle. From \eqs{Neu:Sec1:03a} and (\ref{Neu:Sec1:03b}) from \sect{Neu:Sec:SpinIntro}, along with \eqs{Neu:Sec2:02} to (\ref{Neu:Sec2:06}) in \sect{Neu:Sec:SpinOsc} it is clear that neutrino spin rotates around the second axis. This enables us to construct the effective Hamiltonian for neutrino spin oscillations in the static spherically symmetric spacetime which takes the following form:
\begin{align}\label{Neu:Sec4:01}
H_{\rm{eff}}=\left(\begin{array}{ll}
0 & -i\Omega _{2}\\
i\Omega _{2} & 0
\end{array}\right)
\end{align}
Thus the effective Hamiltonian depends solely on the oscillation frequency $\Omega _{2}$, which vanishes for neutrino moving on a circular orbit (we only consider the high energy limit in the remaining discussion). Then from \eq{Neu:Sec4:01} the neutrino spin oscillation probability after traversing a distance $r$ in time $t=r/c$ turns out to be:
\begin{align}\label{Neu:Sec4:02}
P(t)=\sin ^{2}\left(\Omega _{2}t\right)
\end{align}
Then we can read off the oscillation frequency $\Omega _{2}$ from \eq{Neu:Sec2:06} and then substitute in \eq{Neu:Sec4:02}, which ultimately leads to the following expression for spin oscillation probability of neutrino for the most general static spherically symmetric spacetime in geodesic with energy per particle mass $E$ and corresponding angular momentum $L$ in the high energy limit as:
\begin{align}\label{Neu:Sec4:03}
P(L,E,t)\vert _{\rm{geod}}=\sin ^{2}\left(\left\lbrace \frac{Lf\sqrt{g}}{2Er^{2}}\left[1-\frac{f'r}{2f} \right] \right\rbrace t \right)
\end{align}
The most striking feature of the above expression is that above all other dependencies the oscillation probability depends on the angular momentum $L$. This clearly depicts the fact that if we consider only radial motion, the spin oscillation probability identically vanishes. Thus for \emph{any} static, spherically symmetric spacetime neutrinos traveling along radial direction would \emph{not} suffer helicity flip. If it was left handed originally it will remain left handed for radial motion.  

Note that in all the three cases considered above since $f(r)$ and $g(r)$ are asymptotically flat and there is an additional $1/r^{2}$ factor in the probability. Thus at large distance (for example on earth) the probability for helicity flip will be vanishingly small. Thus in this case we consider two more alternative gravity theories namely, (a) The Einstein-Maxwell-Gauss-Bonnet gravity and (b) $f(R)$ gravity theory both of them are asymptotically de-Sitter or Anti de-Sitter. Thus they can produce significant probability for neutrino helicity flip and we can obtain some bound on their parameters to ensure that the helicity flip remains within experimental bounds. We can also compare these bounds with previously obtained results in \cite{Chakraborty2014CQG,Chakraborty2014PRD8902}. 

\subsection{Einstein-Maxwell-Gauss-Bonnet Gravity}

Spacetime having more than four dimensions is an interesting concept. For it can solve for various fundamental problems in theoretical physics from cosmological constant to hierarchy problem \cite{Lorenzana2005,Chakraborty2014PRD8912,ChakrabortyEPJC2014,Brax2003}. In these extra dimensional scenarios, the spacetime we live in is assumed to be a four dimensional brane, embedded in a higher dimensional bulk. Ordinary matter fields are confined in the brane, while gravity can propagate in the bulk as well. Also as we have argued earlier the Einstein-Hilbert action is supposed to be a low energy realization of the fundamental theory. To keep out the ghost terms it is instructive to modify the gravity action by including the second order Lovelock term which is known in the literature as the Gauss Bonnet term. After inclusion of the Gauss-Bonnet (GB) term and a Maxwell field along with the Einstein-Hilbert term the modified action looks like
\begin{eqnarray}
S&=&\int dx^{5}\sqrt{-g}\Big[R+\alpha \Big(R_{\mu \nu \alpha \beta}R^{\mu \nu \alpha \beta}-4R_{\mu \nu}R^{\mu \nu}
+R^{2}\Big)+F_{\alpha \beta}F^{\alpha \beta} \Big]
\end{eqnarray}
In the above expression $R$, $R_{\mu \nu}$ and $R_{\mu \nu \alpha \beta}$ are Ricci scalar, Ricci tensor and Riemann tensor respectively, $F_{\mu \nu}$ is the electromagnetic field tensor and $\alpha$ being the GB coupling coefficient with dimension of length squared. The field equations for gravity can be obtained by varying the above action with respect to the metric $g_{\alpha \beta}$ and variation of the electromagnetic field tensor $F_{\mu \nu}$ would lead to electromagnetic field equations respectively. Through this variation we obtain \cite{Dehghani04,Chakraborty2009,Dehghani04b,Corradini04}
\begin{eqnarray}
R_{\mu \nu}&-&\frac{1}{2}g_{\mu \nu}R-\alpha \Big[\frac{1}{2}g_{\mu \nu}\Big(R_{\alpha \beta \gamma \delta}R^{\alpha \beta \gamma \delta}
-4R_{\alpha \beta}R^{\alpha \beta}+R^{2}\Big)-2RR_{\mu \nu}+4R_{\mu \alpha}R^{\alpha}_{\nu}
\nonumber
\\
&+&4R^{\alpha \beta}R_{\mu \alpha \nu \beta}-2R_{\mu}^{~\alpha \beta \gamma}R_{\nu \alpha \beta \gamma}\Big]=T_{\mu \nu};\qquad \nabla _{\mu}F^{\mu \nu}=0
\end{eqnarray}
where $T_{\mu \nu}$ is the usual stress tensor for electromagnetic field. The important thing to notice
is that the field equation only contains second order derivatives of the metric, no higher derivatives
are present. This is expected since Gauss-Bonnet gravity is a subclass of Lovelock gravity, which does not contain higher derivative terms of the Riemann tensor.

We can obtain static spherically symmetric solutions to these field equations having the form of Eq. (\ref{Neu:Sec2:01}). It turns out that these solutions are asymptotically de-Sitter or Anti de-Sitter \cite{Dehghani04}. Also we should mention that the solution obtained from the above field equations will be in higher dimensions, i.e., the line element would correspond to: $ds^{2}=-f(r)dt^{2}+f^{-1}(r)dr^{2}+r^{2}d\Omega _{3}^{2}$, where $d\Omega _{3}^{2}=d\theta _{1}^{2}+\sin ^{2}\theta _{1}(d\theta _{2}^{2}+\sin ^{2}\theta _{2}d\theta _{3}^{2})$. However we have been emphasizing that this solution should be interpreted from a brane world point of view, with the visible brane characterized by $\theta _{3}=\textrm{constant}$ hypersurface. With this choice the solution reduces to four dimensions with $(t,r,\theta _{1},\theta _{2})$ as the set of coordinates. Such spherically symmetric solutions were obtained in Ref. \cite{Dehghani04} and has the particular form
with reference to Eq. (\ref{Neu:Sec2:01}) as
\begin{equation}
f(r)=g(r)=K+\frac{r^{2}}{4\alpha}\left[1\pm \sqrt{1+\frac{8\alpha \left(m+2\alpha \mid K \mid \right)}{r^{4}}-\frac{8\alpha q^{2}}{3r^{6}}} \right]
\end{equation}
where $K$ determines the scalar curvature of the spacetime which can take values $0,\pm 1$. However in this work we attribute to $K$ the value 1. Then form solar system tests and neutrino oscillation experiments \cite{Chakraborty2014CQG,Chakraborty2014PRD8902} we can infer stringent bounds on $\alpha^{-1}$. As we have mentioned that we will work at a large distance from the source thus for our study we can make a power series expansion of the terms inside the square root in inverse powers of $r$ and arrive at the following result
\begin{equation}
f(r)=g(r)=1+\frac{r^{2}}{2\alpha}+\frac{m+2\alpha}{r^{2}}-\frac{q^{2}}{3r^{4}}
\end{equation}
Then in the large $r$ limit we have the following expression $rf'/2f\sim 1-(2\alpha /r^{2})$. Using the large $r$ limit of both $f(r)$ and $g(r)$ we obtain the neutrino helicity flip probability to be,
\begin{equation}
P(L,E,\alpha)\vert _{\rm{geod}}=\sin ^{2}\left(\frac{L}{2E\sqrt{2\alpha}}\right)
\end{equation}
\begin{figure*}
\begin{center}

\includegraphics[height=3in, width=5in]{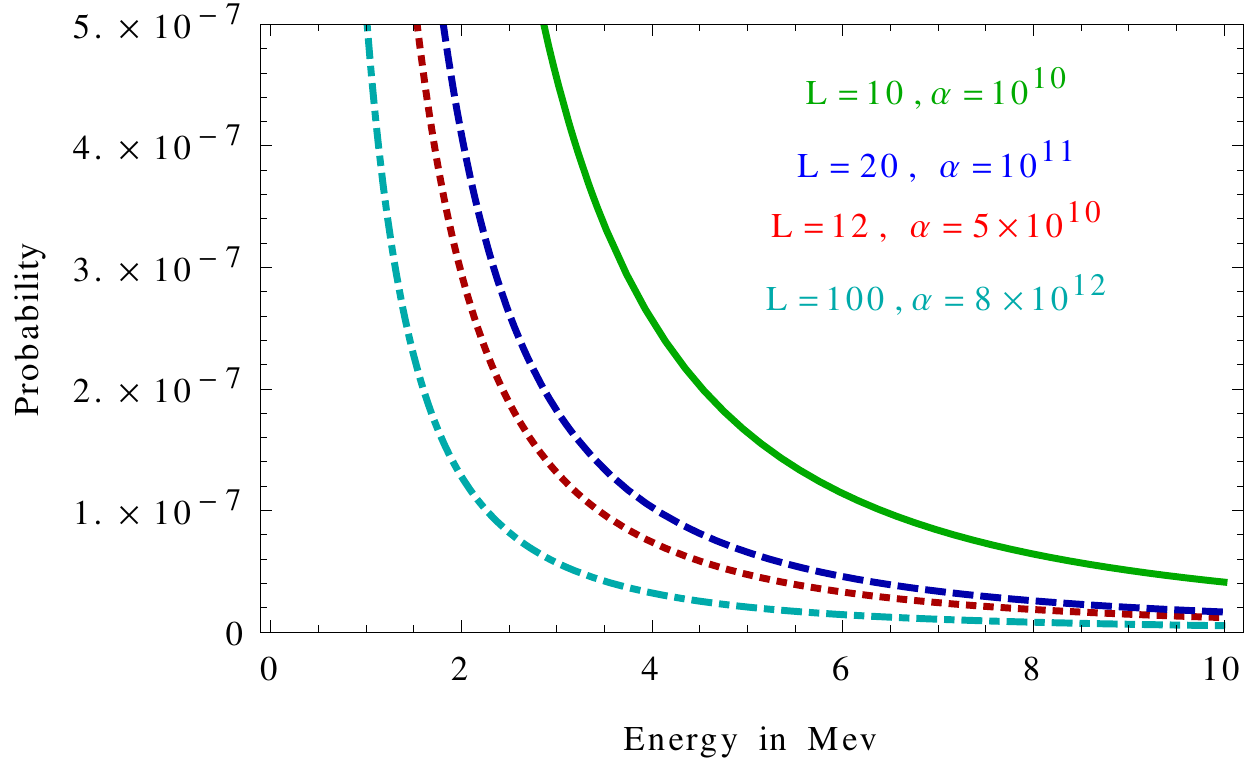}

\caption{(color online) In this figure the variation of neutrino oscillation probability with neutrino energy has been presented for various choices of angular momentum $L$ and the GB parameter $\alpha$. See text for more discussions.}\label{Neu:Fig:07}

\end{center}
\end{figure*}
Note that the helicity flip depends on the angular momentum of the particle $L$, the energy $E$ and the GB coupling parameter $\alpha$. In figure \ref{Neu:Fig:07} we have plotted the helicity flip probability with the energy of the neutrino. It turns out that for all values of $\alpha$ and $L$ the probability of helicity flip is larger for low energy neutrino. While for high energy neutrino the helicity flip probability is substantially small. Also as the GB parameter $\alpha$ increases (or equivalently as $\alpha ^{-1}$ decreases) the probability of neutrino helicity flip decreases significantly. The low energy solar neutrinos have energy of about $0.4$ MeV. Then for $L\sim 100$ and $\alpha \sim 10^{10}$ we get $P\sim 0.02$. This is quite consistent with observations as well \cite{Duan1992}, where from the Kamiokande-II data the helicity flip probability was obtained as $\sim <0.07$. Thus the helicity flip probability of neutrino leads to a bound on the GB parameter which turns out to be $\alpha \sim > 9.34\
times 10^{9}$.
\subsection{f(R) Gravity Theory}

There exist another way of modification of the \EH action, which is obtained by introducing a term $f(R)$ in the \gr\ lagrangian, where $f$ is taken to be some arbitrary function of the scalar curvature $R$. In order to get the gravitational field equations we will use the standard method, i.e., we will vary the metric $g_{\mu \nu}$ leading to \cite{nel10,cor10,Nojiri2011}
\begin{equation}\label{va13}
\frac{1}{2}g_{\mu \nu}f(R)-R_{\mu \nu}f'(R)-g_{\mu \nu}\square f'(R)+\nabla _{\mu}\nabla _{\nu}f'(R) =-4\pi T_{\mu \nu}^{\rm matter}
\end{equation}
The vacuum solution, i.e., solution with $T^{\rm matter}_{ab}=0$ corresponds to a de-Sitter Schwarzschild or Anti de-Sitter Schwarzschild solution for which the Ricci scalar is covariantly constant. This constant Ricci scalar in turn corresponds to  $R_{\mu \nu}\propto g_{\mu \nu}$. Since $\square f'(R)=0$ in the scenario we are considering the field equation given in Eq.~ ($\ref{va13}$) reduces to the following algebraic equation, $0=2f(R)-Rf'(R)$. This immediately suggests that the model $f(R) \propto R^{2}$ satisfy the above equation \cite{Nojiri2011}. Hence the de-Sitter or Anti de-Sitter Schwarzschild solution is an exact vacuum solution the the $f(R)$ gravity theory with the line element being given by (see \eq{Neu:Sec2:01})
\begin{equation}\label{va15}
ds^{2}=\left(1-\frac{2M}{r}\mp \frac{r^{2}}{L_{f}^{2}}\right)dt^{2}- \left(1-\frac{2M}{r}\mp \frac{r^{2}}{L_{f}^{2}}\right)^{-1}dr^{2}-r^{2}d\Omega ^{2}
\end{equation}
which immediately shows that $f(r)=g(r)$. Here the minus(plus) sign corresponds to the spacetime being (Anti) de-Sitter spacetime, $M$ corresponds to the mass of the black hole and $L_{f}$ is the length parameter of the (Anti) de-Sitter spacetime. This length can be related to the scalar curvature as $R=\pm \frac{12}{L_{f}^{2}}$ (where also the plus sign corresponds to de-Sitter spacetime and minus sign corresponds to Anti de-Sitter spacetime). Then in the large $r$ limit we have $rf'/2f\sim 1\pm (L_{f}^{2}/r^{2})$. Using the large $r$ limit of both $f(r)$ and $g(r)$ we obtain the neutrino helicity flip probability to be,
\begin{equation}
P(L,E,L_{f})\vert _{\rm{geod}}=\sin ^{2}\left(\frac{L}{2EL_{f}}\right)
\end{equation}
\begin{figure*}
\begin{center}

\includegraphics[height=3in, width=5in]{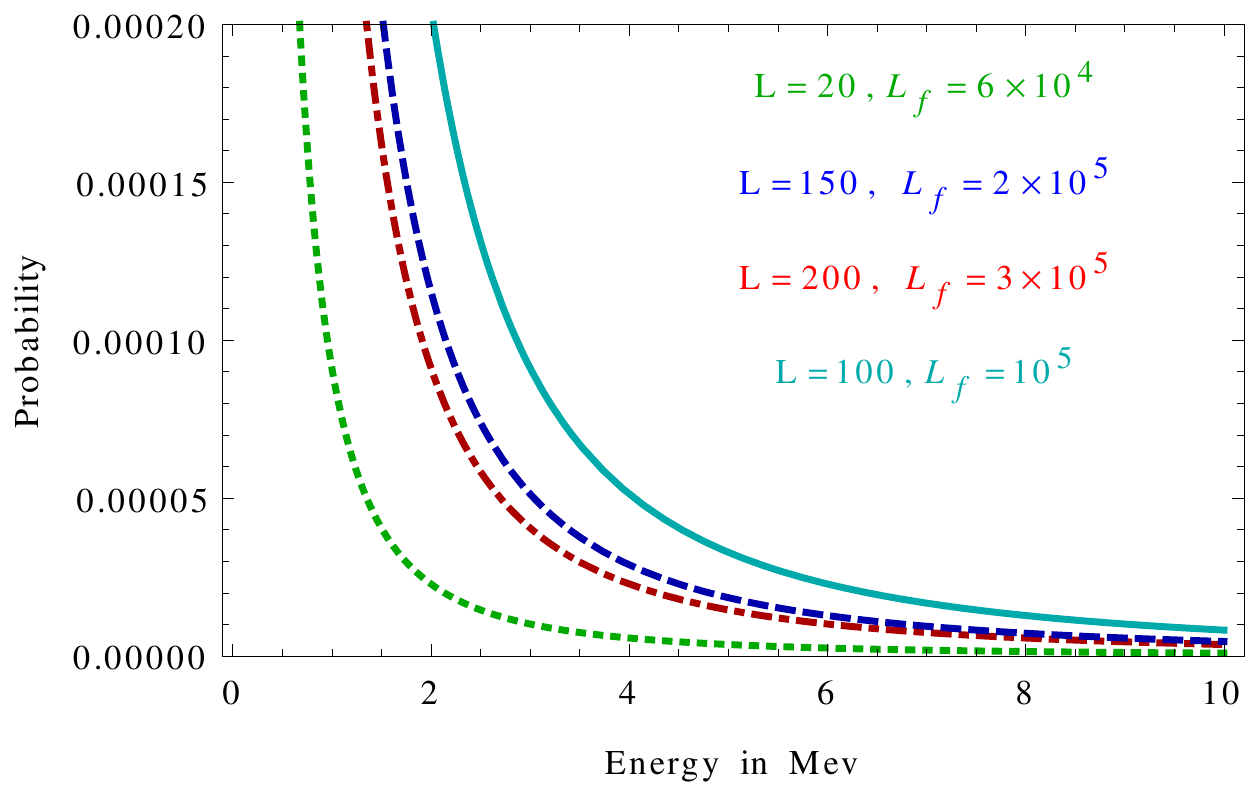}

\caption{(color online) In this figure the variation of neutrino oscillation probability with neutrino energy has been presented for various choices of angular momentum $L$ and the length parameter $L_{f}$ in $f(R)$ gravity model. See text for more discussions.}\label{Neu:Fig:08}

\end{center}
\end{figure*}
The expression for helicity flip probability clearly shows that it depends on the angular momentum of the particle $L$, the energy $E$ and the length parameter $L_{f}$ in $f(R)$ gravity. In figure \ref{Neu:Fig:08} we have plotted the helicity flip probability with the neutrino neutrino energy for different choices of other two parameters. As in the GB scenario in this case as well it turns out that for all values of $L_{f}$ and $L$ the probability of helicity flip is larger for low energy neutrinos. While for high energy neutrinos the helicity flip probability is substantially smaller. Also as the length parameter $L_{f}$ increases the probability of neutrino helicity flip decreases significantly. The low energy solar neutrinos have energy of about $0.4$ MeV. Then for $L\sim 100$ and $L_{f} \sim 10^{5}$ we get $P\sim 0.02$. This is quite consistent with observations as well \cite{Duan1992}, where from the Kamiokande-II data the helicity flip probability was obtained as $\sim <0.07$. Thus the helicity flip 
probability of neutrino leads to a bound on the length parameter in $f(R)$ gravity which turns out to be $L_{f} \sim > 9.87\times 10^{4}$ which is consisten with the bound presented in \cite{Chakraborty2014CQG}.
\section{Discussion}\label{Neu:Sec:Dis}

Neutrino oscillation has been extensively studied in flat spacetime and in general relativistic solutions. Given the recent boost in the search for alternative gravity theories it is legitimate to ask the status of neutrino oscillation in these alternative gravity theories. A first step along this direction was taken in \cite{Chakraborty2014CQG}. In this work we have generalized their setup by considering the most general static spherically symmetric spacetime. Our strategy in this work is as follows: 
\begin{itemize}
\item First we have derived general results for an arbitrary static spherically symmetric spacetime in the context of both spin and flavour oscillation. For neutrino spin oscillation we have shown that the oscillation frequency is dependent on the angular momentum, energy and both $g_{tt}$ and $g_{rr}$ components of the metric. It turns that for \emph{radial} motion the spin oscillation frequency (and hence the probability) identically \emph{vanishes}. Also for circular orbits in the high energy limit the oscillation frequency vanishes. Both these results hold for \emph{any} static spherically symmetric spacetime.
\item For neutrino flavour oscillation we have derived both the oscillation phase and the oscillation length for our general setup. It turns out that the oscillation length depends \emph{only} on the $g_{tt}$ component. Also starting from the oscillation length we have two more quantities, namely departure of the oscillation length from both the flat spacetime and the corresponding \gr solution provides an ideal test to obtain departure of alternative theories from their \gr counterpart.
\item We have discussed three alternative theories to illustrate the results obtained in the context of our general static spherically symmetric spacetime. These three theories include dilaton coupled Maxwell field, \EH action modified with all possible quadratic correction terms and finally regular black hole solutions. By calculating the oscillation probability in all these theories and comparing them with the solar neutrino results we can put bounds on the parameters in these models. It turns out that for dilaton coupled Maxwell field the dilaton charge has the following bound $D<\sim 1.69\times 10^{-9}$ from the SNO Phase-II data. For the quadratic gravity theory the bound on $\alpha _{3}^{-1}$ turns out to be $\alpha _{3}^{-1}<0.43\times 10^{-18}$ obtained from SNO phase-I data. While for regular black holes no such bound was obtained as for them the corrections are within experimental errors.
\item In the case of neutrino spin oscillation, we have considered two separate gravity theory. The $f(R)$ gravity theory and the Gauss-Bonnet gravity theory. In both the theories the helicity flip probability depends on neutrino energy and it turns out to be significant when the neutrino energy is small, while the probability is quite small for high energy neutrinos. This feature was observed in both the theories. Moreover using the helicity flip probability we obtain the following bounds $\alpha \sim >9.34\times 10^{9}$ for Gauss Bonnet parameter and $L_{f}\sim >9.87\times 10^{4}$ for length parameter in vacuum $f(R)$ gravity.  
\end{itemize}
Thus, starting from a general static spherically symmetric metric ansatz we have derived both neutrino flavour oscillation probability and neutrino spin oscillation probability. After obtaining these general results we have applied them to various static spherically symmetric solutions in alternative gravity theories. These in turn when compared with experiments produce experimental bounds on various parameters in these alternative theories.

\section*{Acknowledgements}

Research of S.C. is funded by a SPM fellowship from CSIR, Government of India. The author also thanks the reviewer for useful comments and suggestions.

\appendix
\section{Detailed Expressions for Various Quantities}\label{Neu:AppA}

In this and subsequent sections we provide a detailed analysis of the expressions. We have not included these results in the main text in order to maintain the flow of ideas in the work unhindered. Thus with the view of being helpful to the readers we present the detailed calculation. 

\subsection{Expressions Related to Spin Oscillation Frequency}\label{Neu:AppA:A}

We will start this section with the introduction of the line element, 
\begin{equation}\label{Neu:AppA:01}
ds^{2}=-f(r)dt^{2}+\frac{dr^{2}}{g(r)}+r^{2}d\Omega ^{2}
\end{equation}
Then the Christoffel symbols corresponding to the above metric ansatz are given by:
\begin{align}
\Gamma ^{t}_{rt}&=\frac{f'}{2f}
\nonumber
\\
\Gamma ^{r}_{rr}&=-\frac{g'}{2g};\qquad \Gamma ^{r}_{tt}=\frac{gf'}{2};\qquad 
\Gamma ^{r}_{\theta \theta}=-rg;\qquad \Gamma ^{r}_{\phi \phi}=-rg\sin ^{2}\theta 
\nonumber
\\
\Gamma ^{\theta}_{\phi \phi}&=-\sin \theta \cos \theta ;\qquad \Gamma ^{\theta}_{r\theta}=\frac{1}{r}
\nonumber
\\
\Gamma ^{\phi}_{r\phi}&=\frac{1}{r};\qquad \Gamma ^{\phi}_{\theta \phi}=\cot \theta 
\end{align}
The vierbein vectors have the following expressions:
\begin{align}
e^{(0)}_{\mu}&=\left(\sqrt{f},0,0,0\right);\qquad e^{(1)}_{\mu}=\left(0,\frac{1}{\sqrt{g}},0,0\right)
\nonumber
\\
e^{(2)}_{\mu}&=\left(0,0,r,0\right);\qquad e^{(3)}_{\mu}=\left(0,0,0,r\sin \theta \right)
\end{align}
while the raised components are
\begin{align}
e^{\mu}_{(0)}&=\left(\frac{1}{\sqrt{f}},0,0,0\right);\qquad e^{\mu}_{(1)}=\left(0,\sqrt{g},0,0\right)
\nonumber
\\
e^{\mu}_{(2)}&=\left(0,0,\frac{1}{r},0\right);\qquad e^{\mu}_{(3)}=
\left(0,0,0,\frac{1}{r\sin \theta}\right).
\end{align}
Then we have the following expression for various covariant derivative components of the vierbein vectors as
\begin{align}
\nabla _{\nu}e_{(0)\mu}&=\frac{f'}{2\sqrt{f}}\delta ^{1}_{\mu}\delta ^{0}_{\nu}
\nonumber
\\
\nabla _{t}e_{(1)t}&=-\frac{1}{2}f'\sqrt{g};\qquad \nabla _{\theta} e_{(1)\theta}=r\sqrt{g};
\qquad \nabla _{\phi}e_{(1)\phi}=r\sqrt{g}\sin ^{2}\theta 
\nonumber
\\
\nabla _{\theta}e_{(2)r}&=-1;\qquad \nabla _{\phi}e_{(2)\phi}=r\sin \theta \cos \theta 
\nonumber
\\
\nabla _{\phi}e_{(3)r}&=-\sin \theta ;\qquad \nabla _{\phi}e_{(3)\theta}=-r\cos \theta .
\end{align}
Components of $u^{(a)}=e^{(a)}_{\mu}U^{\mu}$ are given by:
\begin{align}
u^{(0)}=\sqrt{f}U^{0};\qquad u^{(1)}=\frac{1}{\sqrt{g}}U^{1};\qquad u^{(2)}=rU^{\theta};
\qquad u^{(3)}=r\sin \theta U^{\phi}
\end{align}
The components of $\mathcal{G}_{(a)(b)}$ are as follows:
\begin{align}
\mathcal{G}_{(0)(1)}&=\frac{f'}{2}\sqrt{\frac{g}{f}}U^{0};\qquad \mathcal{G}_{(0)(2)}=0=\mathcal{G}_{(0)(3)}
\nonumber
\\
\mathcal{G}_{(1)(0)}&=-\frac{f'}{2}\sqrt{\frac{g}{f}}U^{0};\qquad \mathcal{G}_{(1)(2)}=\sqrt{g}U^{\theta};
\qquad \mathcal{G}_{(1)(3)}=\sqrt{g}\sin \theta U^{\phi}
\nonumber
\\
\mathcal{G}_{(2)(3)}&=\cos \theta U^{\phi}
\end{align}
Hence components of electric and magnetic fields are:
\begin{align}
E_{(1)}&=\frac{f'}{2}\sqrt{\frac{g}{f}}U^{0};\qquad E_{(2)}=E_{(3)}=0
\\
B_{(1)}&=\cos \theta U^{\phi};\qquad B_{(2)}=-\sqrt{g}\sin \theta U^{\phi};\qquad B_{(3)}=\sqrt{g}U^{\theta}
\end{align}
Then we get the components of $G_{a}$ as:
\begin{align}
G_{(1)}&=\frac{1}{2}\cos \theta U^{\phi}+\frac{E^{(2)}u^{(3)}-E^{(3)}u^{(2)}}{2(1+u^{(0)})}
\nonumber
\\
&=\frac{1}{2}\cos \theta U^{\phi}
\\
G_{(2)}&=-\frac{1}{2}\sqrt{g}\sin \theta U^{\phi}+\frac{E^{(3)}u^{(1)}-E^{(1)}u^{(3)}}{2(1+u^{(0)})}
\nonumber
\\
&=-\frac{1}{2}\sqrt{g}\sin \theta U^{\phi}+\frac{f'\sqrt{(g/f)}}{4(1+u^{(0)})}r\sin \theta U^{t}U^{\phi}
\\
G_{(3)}&=\frac{1}{2}\left(\sqrt{g}U^{\theta}-\frac{f'\sqrt{(g/f)}}{2(1+\sqrt{f}U^{t})}u^{(2)}\right)
\nonumber
\\
&=\frac{U^{\theta}}{2}\left(\sqrt{g}-\frac{f'\sqrt{(g/f)}r}{2(1+\sqrt{f}U^{t})}\right)
\end{align}
If we specify to the $\theta =\pi/2$ plane, then we readily arrive at the following expressions for the components of $G_{(\alpha)}$ as, $d\theta /d\tau =0$, and consequently $U^{\theta}=0$ as,
\begin{align}
G_{(1)}&=0;\qquad G_{(3)}=0;
\nonumber
\\
G_{(2)}&=-\frac{U^{\phi}}{2}\left(\sqrt{g}
-\frac{f'\sqrt{(g/f)}}{2(1+\sqrt{f}U^{t})}U^{t}r\right)
\end{align}
These are the expressions used in the main text while calculating the spin oscillation frequency.

\subsection{Expressions Related to Flavour Oscillation Frequency}\label{Neu:AppA:B}

We will study the flavour oscillation frequency for the general metric ansatz as presented in \eq{Neu:AppA:01}. The conserved energy and angular momentum along with radial momentum has the expression:
\begin{equation}
p_{t}^{(k)}=-m_{k}E_{k};\qquad p^{(k)}_{r}=m_{k}\frac{\dot{r}}{g(r)};\qquad p_{\phi}^{(k)}=m_{k}L_{k}
\end{equation}
where, $E$ and $L$, were defined through the relations: $\dot{t}=E/f$ and $\dot{\phi}=L/r^{2}$. Defining the following function:
\begin{equation}
V=1-\frac{fL_{k}^{2}}{r^{2}E_{k}^{2}}
\end{equation}
we arrive at the required expression for $\dot{r}$ given by:
\begin{equation}
\dot{r}=\sqrt{\frac{g}{f}}\sqrt{E_{k}^{2}V-f}
\end{equation}
Then the derivatives $dt/dr$ and $d\phi /dr$ can be obtained as,
\begin{align}
\frac{dt}{dr}&=\frac{\dot{t}}{\dot{r}}=\frac{E_{k}}{\sqrt{fg}\sqrt{E_{k}^{2}V-f}}
\\
\frac{d\phi}{dr}&=\frac{\dot{\phi}}{\dot{r}}=\frac{L_{k}}{r^{2}\sqrt{E_{k}^{2}-V}}\sqrt{\frac{f}{g}}
\end{align}
Finally the upper component of $p^{(k)r}$ has the following expression:
\begin{equation}
p^{(k)r}=m_{k}\sqrt{\frac{g}{f}}\sqrt{E_{k}^{2}V-f}
\end{equation}
Hence the phase along the geodesic of neutrino, with the assumption that it is massive leads to
\begin{align}
\Phi ^{(k)}_{geod}&=-\int _{A}^{B} dr \left[-\frac{m_{k}E_{k}^{2}}{\sqrt{fg}\sqrt{E_{k}^{2}V-f}}
+m_{k}\frac{\sqrt{E_{k}^{2}V-f}}{\sqrt{fg}}
+\frac{m_{k}^{2}L_{k}^{2}}{r^{2}}\sqrt{\frac{f}{g}}\frac{1}{\sqrt{E_{k}^{2}V-f}} \right]
\nonumber
\\
&=-\int _{A}^{B} dr \left[-\frac{m_{k}E_{k}^{2}V}{\sqrt{fg}\sqrt{E_{k}^{2}V-f}}
+m_{k}\frac{\sqrt{E_{k}^{2}V-f}}{\sqrt{fg}}\right]
\nonumber
\\
&=\int _{A}^{B} dr \left[\frac{m_{k}}{\sqrt{E_{k}^{2}V-f}}\sqrt{\frac{f}{g}}\right]
\end{align}
However in the literature sometimes the neutrino though have a mass is taken to travel along the null geodesics, as it is extremely relativistic. For that purpose the expressions for $\dot{r}$, $dt/dr$ and $d\phi/dr$ leads to,
\begin{equation}
\dot{r}=E_{k}\sqrt{V}\sqrt{\frac{g}{f}};\qquad \frac{dt}{dr}=\frac{1}{\sqrt{gfV}};\qquad 
\frac{d\phi}{dr}=\frac{L_{k}}{E_{k}r^{2}\sqrt{V}}\sqrt{\frac{f}{g}}
\end{equation}
Then the phase along null trajectory can be presented as:
\begin{align}
\Phi ^{(k)}_{null}&=-\int ^{B}_{A}dr \left[-\frac{m_{k}E_{k}}{\sqrt{gfV}}
+\frac{m_{k}L_{k}^{2}}{E_{k}r^{2}\sqrt{V}}\sqrt{\frac{f}{g}}+\frac{m_{k}}{\sqrt{fg}}\sqrt{E_{k}^{2}V-f}
\right]
\nonumber
\\
&=-\int ^{B}_{A}dr \left[-\frac{m_{k}E_{k}\sqrt{V}}{\sqrt{gf}}
+\frac{m_{k}\sqrt{E_{k}^{2}V-f}}{\sqrt{fg}}\right]
\nonumber
\\
&=\int ^{B}_{A}dr \left[\frac{m_{k}}{2E_{k}\sqrt{V}}\sqrt{\frac{f}{g}} \right]
\end{align}

\section{Circular Orbit in Spherically Symmetric Spacetime}\label{Neu:AppB}

We start with the Lagrangian on the equatorial i.e. $\theta =\pi/2$ plane, by exploiting the spherical symmetry of the problem, such that
\begin{equation}
L=-\frac{1}{2}f \left(\frac{dt}{d\tau}\right)^{2}+\frac{1}{2g}\left(\frac{dr}{d\tau}\right)^{2}
+\frac{1}{2}r^{2}\left(\frac{d\phi}{d\tau}\right)^{2}
\end{equation}
Then the energy and the angular momentum are conserved, since the Lagrangian does not involve any functions of time or of azimuthal angle $\phi$ \cite{Chakraborty2011CJP}. Then we get the equation of motion for $\dot{t}$ and $\dot{\phi}$ as:
\begin{align}
\dot{t}&=\frac{E}{f};\qquad \left(\textrm{since} \frac{d}{d\tau}\left(f\dot{t}\right)=0 \right)
\\
\dot{\phi}&=\frac{L}{r^{2}};\qquad \left(\textrm{since}\frac{d}{d\tau}\left(L\dot{\phi}\right)=0 \right)
\end{align}
From the above expressions it is evident that the four-velocity components are being given by:
\begin{equation}
U^{\mu}=\left(\frac{E}{f},\frac{dr}{d\tau},0,\frac{L}{r^{2}} \right)
\end{equation}
Thus using the on-mass shell condition: $p_{\mu}p^{\mu}=-1$, we arrive at the equation for $\dot{r}$ leading to the following expression:
\begin{equation}
\left(\frac{dr}{d\tau}\right)^{2}=\frac{g}{f}\left[E^{2}-f\left(1+\frac{L^{2}}{r^{2}} \right) \right]
\end{equation}
Then along with this relation we can use the expression for $\dot{\phi}$ leading to the equation for the orbit of a massive particle as:
\begin{equation}
\left(\dfrac{dr}{d\phi}\right)^{2}=\frac{g(r)r^{4}}{f(r)L^{2}}
\left[E^{2}-f\left(1+\frac{L^{2}}{r^{2}} \right)  \right]
\end{equation}
Then introducing a new variable $r=(1/u)$, we get, $(dr/d\phi)=-(1/u^{2})(du/d\phi)$, which modifies the above equation to the form:
\begin{equation}
\left(\dfrac{du}{d\phi}\right)^{2}=\frac{g(r)}{f(r)L^{2}}
\left[E^{2}-f\left(1+\frac{L^{2}}{r^{2}} \right)  \right]
\end{equation}
Differentiating this expression again we obtain the following second order differential equation satisfied by the variable $u$ as:
\begin{equation}
\frac{d^{2}u}{d\phi ^{2}}=g\left[-u+\frac{g'}{2g}\left(\frac{1+L^{2}u^{2}}{L^{2}u^{2}}\right)
-\frac{E^{2}}{2L^{2}u^{2}}\left(\frac{g'}{gf}-\frac{f'}{f^{2}} \right) \right]
\end{equation}
where ``prime" denotes derivative with respect to the radial coordinate $r$. For circular orbit, $u$ is fixed, say at $u=u_{0}$. Then $(d^{2}u/d\phi ^{2})$ should vanish along with $(dr/d\phi)$. This leads to the following equations:
\begin{align}
E^{2}\left(\frac{g'}{f}-\frac{gf'}{f^{2}}\right)&+L^{2}\left(2gu_{0}^{3}-g'u_{0}^{2}\right)=g'
\\
E^{2}\frac{g}{f}-gL^{2}u_{0}^{2}&=g
\end{align}
Then the energy and angular momentum for the circular orbit are obtained as:
\begin{align}
E_{c}&=\sqrt{\frac{2f^{2}}{2f-r_{0}f'}}
\\
L_{c}&=\sqrt{\frac{r^{3}f'}{2f-r_{0}f'}}
\end{align}
Thus the four-velocity components are being given by:
\begin{equation}
U^{\mu}=\left(\sqrt{\frac{2}{2f-r_{0}f'}},0,0,
\sqrt{\frac{f'}{r_{0}\left(2f-r_{0}f'\right)}}\right)
\end{equation}
Thus we get the following angular velocity having the expression:
\begin{equation}
\frac{d\phi}{dt}=\frac{U^{\phi}}{U^{t}}=\sqrt{\frac{f'}{2r_{0}}}
\end{equation}
However for null trajectory i.e. for photons we get equation for orbit as:
\begin{equation}
\left(\frac{dr}{d\phi}\right)^{2}=\frac{r^{4}g}{L^{2}}\left(\frac{E^{2}}{f}-\frac{L^{2}}{r^{2}}\right)
\end{equation}
the above equation for the orbit can be written introducing the new variable $u=(1/r)$ to have the following form:
\begin{equation}
\frac{d^{2}u}{d\phi ^{2}}=g\left[-u+\frac{g'}{2g}-\frac{E^{2}}{2L^{2}u^{2}}
\left(\frac{g'}{gf}-\frac{f'}{f^{2}}\right) \right]
\end{equation}
Then from the condition of circular orbit at $r=r_{0}$, or equivalently $u=u_{0}$, we get the following relations between energy and angular momentum such that:
\begin{align}
\frac{E^{2}}{L^{2}}&=\frac{f(r_{0})}{r_{0}^{2}}
\label{Neu:AppB:17}
\\
u_{0}&=\frac{f'(r_{0})}{2f(r_{0})}
\label{Neu:AppB:18}
\end{align}
For Schwarzschild spacetime, we get the following equation: $2Mu_{0}^{2}=2u_{0}(1-2Mu_{0})$, leading to $r_{0}=3M$ as the photon circular orbit radius. Then four velocity has the expression:
\begin{equation}
U^{\mu}=\left(\frac{E}{f},0,0,\frac{E}{r_{0}\sqrt{f}}\right)
\end{equation}
Thus the corresponding angular velocity turns out to be:
\begin{equation}
\frac{d\phi}{dt}=\frac{U^{\phi}}{U^{t}}=\frac{1}{r_{0}}\sqrt{f}
\end{equation}
The above expressions for energy and angular momentum, along with the angular velocity have been used extensively in the main text.



\end{document}